\documentclass[aps,prb,twocolumn,amsmath,amssymb,superscriptaddress]{revtex4}

\usepackage{amssymb}
\usepackage{amsmath}
\usepackage{graphicx}
\usepackage{textcomp}
\usepackage{color}
\usepackage{amsfonts}
\usepackage{epsfig}
\usepackage{hyperref}
\usepackage{bm}
\usepackage[caption=false]{subfig} 
\usepackage{booktabs}
\usepackage{tabularx}
\usepackage{array}

\newcommand{\arctanh}[1]{\operatorname{arctan}}

\newcolumntype{M}[1]{>{\centering\arraybackslash}m{#1}}

\DeclareMathAlphabet\mathbfcal{OMS}{cmsy}{b}{n}

\begin{document}

\title{Spin-orbit induced equilibrium spin currents in materials}
\author{Andrea Droghetti}
\email{andrea.droghetti@tcd.ie}
\affiliation{School of Physics and CRANN, Trinity College, Dublin 2, Ireland}
\author{Ivan Rungger}
\email{ivan.rungger@npl.co.uk}
\affiliation{National Physical Laboratory, Hampton Road, Teddington TW11 0LW, United Kingdom}
\author{Angel Rubio}
\email{angel.rubio@mpsd.mpg.de}
\affiliation{Max Planck Institute for the Structure and Dynamics of Matter, Center for Free-Electron Laser Science and Department of Physics, Luruper Chaussee 149, 22761 Hamburg, Germany}
\affiliation{Center for Computational Quantum Physics, Flatiron Institute, New York, New York 10010, United States}
\affiliation{Nano-Bio Spectroscopy Group and European Theoretical Spectroscopy Facility (ETSF), Departamento de Pol\'imeros y Materiales Avanzados: F\'isica, Qu\'imica y Tecnolog\'ia, Universidad del Pa\'is Vasco (UPV/EHU), Av. Tolosa 72, 20018 San Sebasti\'{a}n, Spain}
\affiliation{Donostia International Physics Center (DIPC), 20018 Donostia-San Sebasti\'{a}n, Spain}
\author{Ilya V. Tokatly}
\email{ilya.tokatly@ehu.es}
\affiliation{Nano-Bio Spectroscopy Group and European Theoretical Spectroscopy Facility (ETSF), Departamento de Pol\'imeros y Materiales Avanzados: F\'isica, Qu\'imica y Tecnolog\'ia, Universidad del Pa\'is Vasco (UPV/EHU), Av. Tolosa 72, 20018 San Sebasti\'{a}n, Spain}
\affiliation{Donostia International Physics Center (DIPC), 20018 Donostia-San Sebasti\'{a}n, Spain}
\affiliation{IKERBASQUE, Basque Foundation for Science, 48009 Bilbao, Spain}
\affiliation{ITMO University, Department of Physics and Engineering, Saint-Petersburg, Russia}

\begin{abstract}
The existence of pure spin-currents in absence of any driving external field is commonly considered an exotic phenomenon appearing only in quantum materials, such as topological insulators.
We demonstrate instead that equilibrium spin currents are a rather general property of materials with non negligible spin-orbit coupling (SOC). Equilibrium spin currents can be present at the surfaces of a slab. Yet, we also propose the existence of global equilibrium spin currents, which are net bulk spin-currents
along specific crystallographic directions of solid state materials. Equilibrium spin currents are allowed by symmetry in a very broad class of systems having gyrotropic point groups. 
The physics behind equilibrium spin currents is uncovered by making an analogy between electronic systems with SOC and non-Abelian gauge
theories. The electron spin can be seen as the analogous of the color degree of freedom in SU$(2)$ gauge theories and equilibrium spin currents can then be identified with diamagnetic color currents appearing as the response to a effective non-Abelian magnetic field generated by the SOC. 
Equilibrium spin currents are not associated with spin transport and accumulation, 
but they should nonetheless be carefully taken into account when computing transport spin currents. 
We provide quantitative estimates of equilibrium spin currents for a number of different systems, 
specifically the Au(111) and Ag(111) metallic surfaces presenting Rashba-like surface states, 
nitride semiconducting nanostructures and bulk materials, such as the prototypical gyrotropic medium tellurium.
In doing so, we also point out the limitations of model approaches showing that first-principles calculations are needed
to obtain reliable predictions. We therefore use Density Functional Theory
computing the so-called bond currents, which represent a powerful tool to deeply understand the relation between equilibrium currents, electronic structure 
and crystal point group. 

\end{abstract}

\maketitle

\section{Introduction}
The spin-orbit coupling (SOC) is one of the most important interactions in spintronics, since it allows to control the spin degree of freedom by electrical means \cite{Awschalom,Jungwirth,Sinova}.
Despite the rapidly growing number of studies dedicated to SOC-driven phenomena, 
some fundamental questions remain debated.
Among these there is the possibility for SOC to induce spin-currents in thermodynamic equilibrium \cite{Rashba,Shi,Sonin1,Sonin2,Tokatly}. \\
Similar to the charge current, which is the flow of electronic charges, the spin current is generally viewed as the flow of angular momentum mediated by electrons and driven by an external stimulus, such as a precessing magnetic field.
According to this picture, no spin current would be expected in absence of any external stimuli and in equilibrium. However equilibrium spin currents (ESCs) are easily computed for the 2D Rashba electron gas and other spin-orbit coupled model systems\cite{Rashba}, including model graphene\cite{Bolivar,HZhang}. This rises questions about the existence of ESCs and their physical interpretation. Part of the controversy comes from the fact that spin, unlike charge, is not a conserved quantity in presence of SOC. Mathematically, the time derivative of the spin density does not reduce to a divergence of a current, but always contains an extra term, the spin-torque \cite{Maekawa_book}. There is therefore an apparent ambiguity in the definition of spin current. Besides this ambiguity, we note that ESCs have so far been computed only in some model systems and one may ask whether they are specific features of those models. In fact, to our knowledge, the existence and eventual magnitude of ESCs in material compounds have never been investigated. In this paper we address these issues.\\ 
The physics behind ESCs is uncovered \cite{Tokatly} by making an analogy between electronic systems with SOC and non-Abelian gauge
theories\cite{Mineev,Frohlich,Jin,Tokatly} using ideas and techniques, typically introduced in the context of quantum chromodynamics. Many different aspects of SOC-related physics in materials, both in the equilibrium and in transport regimes, acquire a simple and natural explanation when SOC is interpreted in terms of an effective non-Abelian SU(2) gauge field\cite{Rebei,Bernevig,Hatano,Yang,Liu,Tokatly2010-1,Tokatly2010-2,Berche,Gorini,Raimondi,Bergeret2014,Bergeret2015}. In this paradigm ESCs are identified with diamagnetic color currents appearing as the response to such field
and aiming at compensating it. This picture is analogous to Landau diamagnetism and it can be seen as its non-Abelian
generalization. ESCs are therefore expected to be present in
almost any physical system. \\
By using first-principles Density Functional Theory (DFT) calculations 
we demonstrate that ESCs emerge in materials whenever 
allowed by crystal symmetry, and that they are found in metals and insulators alike.
In particular, we predict that a global spin current, which is a net ESC along a specific crystallographic direction, 
is present in certain non-centrosymmetric crystals, called gyrotropic media. Since more than half of the crystal point groups are gyrotropic,
ESCs are common and not at all unique to quantum materials, such as topological insulators\cite{Han}. 
Furthermore ESCs are ubiquitous at surfaces and interfaces, where they can be used as a ``measure'' for the effective surface SOC strength. 
We will discuss these main outcomes of our work with several examples pointing out also the limitation of model approaches 
and the need for accurate first-principles calculations to obtain reliable predictions.\\
The paper is organized as follows.
In Section I we consider a general electronic Hamiltonian with SOC and make a link to non-Abelian gauge theories. We then define a gauge invariant spin current and discuss the physical significance of ESCs. 
In Section II we explain how spin-currents can be easily obtained via DFT calculations. In Section III we present examples of ESCs in several materials. 
In particular, we first consider the Au(111) surface, which is a very instructive system  
because of its Rashba-like surface bands. 
We systematically compare ESCs obtained within model descriptions to those computed by means of DFT revealing the relative importance of surface and bulk bands. 
Afterwards we discuss how ESCs emerge in semiconducting nanostructures as well as bulk materials 
and we estimate ESCs in the prototypical gyrotropic material tellurium. Finally we conclude in Section III.
 
\section{Physical interpretation of equilibrium spin currents }\label{sec.spin_currents}
The Hamiltonian of an electron including spin-dependent relativistic corrections up to the order $1/c^2$ (with $c$ the speed of light) is 
\begin{equation}
\hat H=\frac {\hat {\mathbf p}^2}{2m}+ \hat U(\mathbf{r})+\frac{e \hbar}{4m^2 c^2}\hat{\mathbf p}\cdot [\hat{\mathbf\sigma}\times \hat{\mathbf{E}}(\mathbf{r})]+ \frac{ g \mu_B }{2 }\mathbf{B}(\mathbf{r}) \cdot\hat{\mathbf\sigma},\label{general_H}
\end{equation}
and has spinor wave-functions $\Psi^\dagger=\vert \psi^\dagger_\uparrow\,\,\,\psi^\dagger_\downarrow \vert$; $e$ and $m$ are the electron charge and mass;
$\hat{\mathbf r}$, $\hat{\mathbf p}$ and $\hbar\hat{\mathbf\sigma}/2$ are the position, momentum and spin operators; 
$\hat U(\mathbf{r})$ is the scalar external potential. 
The last term is the Zeeman interaction between the spin and an external magnetic field $\mathbf{B}(\mathbf{r})$ with $\mu_B$ the Bohr magneton and $g$ the electron $g$-factor. 
$\mathbf{E}(\mathbf{r})$ is the electric field produced by nuclei in molecules or solids. The third term of $\hat H$ is the SOC interaction term, which up to a prefactor, can be rewritten as
\begin{equation}
\hat{\mathbf p}\cdot [\hat{\mathbf\sigma}\times \hat{\mathbf{E}}(\mathbf{r})]=[\hat{\mathbf{E}}(\mathbf{r})\times \hat{\mathbf p}]\cdot\hat{\mathbf\sigma}=
\frac{E(\mathbf{r})}{r}(\hat{\mathbf{r}}\times \hat{\mathbf p})\cdot\hat{\mathbf\sigma}=\frac{E(\mathbf{r})}{r}\hat{\mathbf{L}}\cdot\hat{\mathbf{\sigma}},
\end{equation}
where $\hat{\mathbf{L}}= \hat{\mathbf{r}}\times \hat{\mathbf p}$ is orbital angular momentum. \\
The problem related to the existence of ESCs is conveniently treated by making a connection to non-Abelian gauge theories
\cite{Mineev,Frohlich,Jin,Tokatly,Berche} and
interpreting the SOC and the Zeeman interaction in Eq. (\ref{general_H}) in terms of a non-Abelian SU(2) vector potential 
$\hat{\mathbfcal{A}}_\mu=\mathcal{A}^a_\mu\hbar\hat{\sigma}^a/2$, where $\mu=0,x,y,z$. 
Specifically, the components of such vector potential read 
\begin{equation}
 \hat{\mathbfcal{A}}_0(\mathbf{r})=-\frac{g\mu_B}{2}  B^a(\mathbf{r})\hat{\sigma}^a \,,\,\,\,\, \hat{\mathbfcal{A}}_i(\mathbf{r})=\frac{e \hbar}{2m c^2}\epsilon_{ija}E_j(\mathbf{r})\hat{\sigma}^a,\label{gauge_field}
\end{equation}
 where $a=x,y,z$ and $i=x,y,z$ respectively label the spin and spatial components
(note that we use Einsten's summation convention on repeated indexes throughout this paper).
The electron Hamiltonian can then be rewritten as
\begin{equation}
 \hat H=\frac {[\hat { p}_i - \hat{\mathbfcal{A}}_i(\mathbf{r})]^2}{2m}-\hat{\mathbfcal{A}}_0(\mathbf{r})   + \hat U(\mathbf{r}),
\end{equation}
where we absorbed the quadratic term $-\hat{\mathbfcal{A}}_i\hat{\mathbfcal{A}}_i/2m$ into the scalar potential. 
The beauty of this representation is that $\hat H$ immediately appears invariant with respect to local non-Abelian gauge transformations
\begin{equation}
\hat{\mathbfcal{A}}_\mu \rightarrow \hat{\mathcal{U}}\hat{\mathbfcal{A}}_\mu 
\hat{\mathcal{U}}^{-1}-i\hbar(\partial_\mu 
\hat{\mathcal{U}})\hat{\mathcal{U}}^{-1},
\end{equation}
where $\hat{\mathcal{U}}(\mathbf{r})=e^{i\eta^a(\mathbf{r})\hat{\sigma}^a/2}$ is an arbitrary 
SU(2) matrix that transform the wave-function as $\Psi\rightarrow \hat{\mathbfcal{U}}\Psi$ (note 
that from now on we will not indicate the explicit dependence on $\mathbf{r}$, unless strictly 
needed, to keep the notation lighter).
The gauge invariance then implies covariant conservation of a current $\mathbf{j}_{\mu}$ \cite{Tokatly}
\begin{equation}
D_0 j_0^a+D_i j_i^a=0,\label{covariant_conservation}
\end{equation}
where 
$D_\mu \cdot= \partial_\mu \cdot -i [\hat {\mathbfcal{A}}_\mu,\cdot]/\hbar$ 
is the covariant derivative and 
\begin{eqnarray}
&j^a_0=\frac{\hbar}{2}\Psi^\dagger(\mathbf{r})\hat\sigma^a\Psi= s^a\\
&j^a_i=\frac{\hbar}{4m}[  \Psi^\dagger \hat\sigma^a (\hat{ p}_i \Psi)+( 
\hat{p}_i\Psi)^\dagger\hat\sigma^a\Psi] 
-\frac{\hbar^2}{4m}\Psi^\dagger\mathcal{A}^a_i\Psi.\label{spin_current}
\end{eqnarray}
Eq. (\ref{covariant_conservation}) is mathematically identical to the covariant conservation of the color degree of freedom associated to the SU(2) quark matter. However, physically, it represents the spin continuity equation, where 
$j^a_0$ is the $a$ component of the spin and $\mathbf{j}^a$ is the corresponding spin current density.
This reasoning is completely analogous to that employed in the familiar case of U(1) gauge fields, where the gauge invariance leads to the charge continuity equation. We note that the definition of spin current density in Eq. (\ref{spin_current}) coincides with the ``common'' definition used by Rashba to predict ESCs in spin-orbit coupled model systems \cite{Rashba}. It can be rewritten as the expectation value $j^a_i=\hbar/4\langle \{\hat \sigma^a,\hat v_i\}\rangle $, where $\hat{v}_i=(i\hbar)^{-1}[\hat{r}_i,\hat H]$ is the $i$-th component of the velocity operator. The last term in Eq. (\ref{spin_current}) owns to the SOC, which introduces a spin-dependent component in the particle velocity, the so-called anomalous term. \\
Explicitly, Eq. (\ref{covariant_conservation}) reads
\begin{equation}
 \partial_t s^a + \epsilon^{abc}\mathcal{A}^b_0 s^c + \partial_i j^a_i + 
 \epsilon^{abc}\mathcal{A}^b_i j^c_i=0. \label{covariant_conservation2} 
\end{equation}
The first term is the rate of change of the spin density at a point in space, while the third term is the divergence of the spin current density. These are analogous to the rate of change of the charge density and to the divergence of the charge current density in the standard U(1) case. However we note that, in SU(2), there are two extra terms in the continuity equation, namely the second and the forth terms in Eq. (\ref{covariant_conservation2}). They express the fact that the spin, unlike the charge, is not conserved. They are the torque caused by the Zeeman magnetic field and by the SOC, respectively. As mentioned in the introduction, the
separation between spin current density and spin-torque in Eq. (\ref{covariant_conservation2}) has been seen as a source of ambiguity in the definition of spin-currents (see for example Ref. \cite{Shi}). 
However, the derivation based on the gauge invariance leaves no room for such ambiguity. It rigorously defines the spin current density according to Eq. (\ref{spin_current}). We are then forced to accept the consequences that follow.
Among these there is the existence of ESCs.\\
In the case of the electromagnetic U(1) gauge field a dissipative charge current is induced by the electric component of the field, 
while non-dissipative diamagnetic currents emerge as the response to the magnetic field. They are calculated as the derivative of the energy with respect to the magnetic vector potential.
In a very same way, in the SU(2) case, dissipative currents are driven by the effective SU(2) 
electric field 
$\hat{\mathbfcal{F}}_{i0}=\partial_i\hat{\mathbfcal{A}}_0-\partial_t\hat{\mathbfcal{A}}_i-i[\hat{
\mathbfcal{A}}_i,\hat{\mathbfcal{A}}_0]/red{\hbar}$,
while there are also non-dissipative currents due to the effective magnetic field 
$\hat{\mathbfcal{F}}_{ij}=\partial_i\hat{\mathbfcal{A}}_j-\partial_j\hat{\mathbfcal{A}}_i-i[\hat{
\mathbfcal{A}}_i,\hat{\mathbfcal{A}}_j]/\hbar$.
In other words, SOC enters the electronic Hamiltonian as an effective background non-Abelian field, and, if a magnetic part of this color field is nonzero, one naturally expects an orbital response in the form of color diamagnetic currents.
The components of the current density are given by the derivative of the energy 
with respect to $\mathcal{A}_i^a$ 
\begin{equation}
 j_{i}^{a}=\langle \delta H/\delta\mathcal{A}_{i}^{a}\rangle\label{eq.DHDA}
\end{equation}
as shown in Ref. \cite{Tokatly} or they can equivalently be calculated from the definition, 
Eq. (\ref{spin_current}), taking the thermodynamic average\cite{Rashba}.
Importantly, since ESCs are non-dissipative currents, they do not transport spin and they do not result in spin accumulation.\\
ESCs can be readily analysed for the models with linear SOC of
Rashba-Dresselhaus form\cite{Rashba}. The only non-zero components of the SU(2) vector potential are 
\begin{equation}
 \hat{\mathbfcal{A}}_x=m(\lambda_D\hat{\sigma}^x- \lambda_R \hat{\sigma}^y)\,,\,\,\,\, \hat{\mathbfcal{A}}_y=m(\lambda_R \hat{\sigma}^x-\lambda_D\hat{\sigma}^y)
\end{equation}
 where $\lambda^{R}$ and $\lambda^{D}$ are the Rashba and Dresselhaus SOC constants. Hence, the
corresponding non-zero components of the ESC density are\cite{Rashba,Tokatly}
 \begin{eqnarray}
 &j^x_x=-j^y_y=\frac{2 \pi}{3} \frac{m^2\lambda_D(\lambda_R^2-\lambda_D^2)}{ \hbar^4 V_{BZ}},\label{SC_xy_Rasbha_Dresselhaus_tot1}\\
 &j_y^x=-j^y_x=\frac{2 \pi}{3} \frac{m^2\lambda_R(\lambda_R^2-\lambda_D^2)}{ \hbar^4 V_{BZ}},\label{SC_xy_Rasbha_Dresselhaus_tot2} 
\end{eqnarray}
where $V_{BZ}$ is the Brillouin zone volume. Notably, $j^x_x$ and $j_y^x$ vanish for the special values $\lambda_R=\pm \lambda_D$. 
The reason is that the color magnetic field $\mathcal{F}_{xy}^z=m^{2}(\lambda_R^2-\lambda_D^2)$ vanishes for this special case. In the absence of any magnetic field there are no diamagnetic currents. 
We therefore see that the physical reason for the equilibrium spin currents is a response to the SOC-induced non-Abelian magnetic field.\\
In spite of the analogy between Landau diamagnetic currents and ESCs, these last ones are somehow more universal. 
Diamagnetic charge currents in a sample are confined at its surface 
and {\it global} equilibrium charge currents, that are net currents through a whole sample cross-section,
are forbidden. For instance, in a finite slab, there will be charge currents at the top and bottom surfaces flowing in opposite directions and therefore compensating,
while the charge current will vanish in the bulk.
This is due to the Bloch-Bohm theorem\cite{Bohm}, which states the impossibility of persistent charge currents in the ground state of an electronic system with a normalizable ground state wave-function.
In contrast, no similar theorem exists for the spin case. 
Global ESCs therefore appear whenever allowed by symmetry. \\
The spin current density is a second-rank pseudotensor, even under time-reversal, because it
transforms as the direct product of the momentum vector and of the spin pseudovector according to Eq. (\ref{spin_current}).
Second-rank pseudotensors are allowed by symmetry only in a subset of non-centrosymmetric systems, called gyrotropic\cite{Ivchenko,Ganichev4}. Gyrotropic materials were first studied because of their optical activity, which is in fact expressed in terms of the second-rank gyration pseudotensor \cite{Landau}.
Of the 32 crystal point groups, 21 are non-centrosymmetric. Among these, $18$ are gyrotropic. The three non-centrosymmetric classes, which are non-gyrotropic, are $\mathbb{T}_d$, $\mathbb{C}_{3h}$, $\mathbb{D}_{3h}$. 
Materials with these three point groups are not expected to show ESCs in spite of being non-centrosymmetric. 
The gyrotropic point groups are $\mathbb{O}$, $\mathbb{T}$, $\mathbb{C}_1$, $\mathbb{C}_2$, $\mathbb{C}_3$, $\mathbb{C}_4$, $\mathbb{C}_6$, $\mathbb{D}_2$, $\mathbb{D}_3$, $\mathbb{D}_4$, $\mathbb{D}_6$, 
$\mathbb{C}_s$, $\mathbb{C}_{2v}$, $\mathbb{C}_{3v}$, $\mathbb{S}_4$, $\mathbb{D}_{2d}$, $\mathbb{C}_{4v}$ and $\mathbb{C}_{6v}$. 
Global ESCs are expected in all compounds with these point symmetries, provided that they have non-negligible SOC. 
As such, ESCs are quite common intrinsic features of materials. This is an important ``take-home'' message of our paper. \\
In centrosymmetric materials, the symmetry can be reduced to gyroptropic, thus leading to the emergence of ESCs, for example through the application of a strain-gradient\cite{Zubko}. Moreover, the inversion symmetry is naturally broken at surfaces and interfaces, which, in most cases, turn out to have gyrotropic point groups. 
ESCs are therefore present at surfaces even in materials, where bulk ESCs are forbidden. As real material samples always have surfaces, ESCs are truly ubiquitous in nature. We will present several illustrative examples in the following sections.\\
The emergence of surface ESCs belongs to the plethora of interfacial phenomena described in terms of the so-called effective interfacial SOC, that is the combination of the atomic SOC with the loss of inversion symmetry. Other well-known examples include spin-charge conversion\cite{Sanchez,Isasa,Karube,Baek} and interfacial spin-orbit torque\cite{Miron,Zhu,Amin}. 
While it is generally assumed that the effective interfacial SOC and the magnitude of interfacial phenomena is exclusively dictated by spin-textured surface bands\cite{Sanchez, Sangiao, Isasa,Chen,Rousseau,XChen}, an analysis based on the ESCs reveals that this is not the case. In sec. \ref{sec.Au}
we will show that surface bands have a minor contribution, while we argue that ESCs are mostly associated to bulk states scattering off the surface. 
Notably, a similar conclusion was reached also by studying current-induced spin polarization at metallic surfaces\cite{Tokatly2,in_preparation}. 
We therefore propose that the calculation of the surface ESCs allows for a practical estimate of effective interfacial SOC in any system, metallic or insulating, with and without surface bands. This is another important message of our work.\\
Finally, we would like to address whether the existence of ESCs can be detected, although they do not lead to any spin accumulation. Different experiments have been proposed. For instance, Sonin suggested to exploit a magneto mechanical effect\cite{Sonin2}. 
If a Rashba 2D medium is integrated into a mechanical cantilever magnetometer, one might be able to measure a mechanical torque.
The argument is based on the observation that  
the ESCs are constant in the bulk of the Rashba medium according to Eqs. (\ref{SC_xy_Rasbha_Dresselhaus_tot1}) and (\ref{SC_xy_Rasbha_Dresselhaus_tot2}) (with $\lambda_D=0$), whereas they must vanish at the very edge. This would lead to an edge orbital torque and to a flux of the orbital moment with a sign opposite to that of the spin thus complying the total angular
momentum conservation law. However, since the Rashba medium has no orbital moment in its 2D plane, the whole
orbital torque must be applied to the free edge of the cantilever, which is then deformed. The idea is intriguing and we note that the required mechanical cantilever was recently realized\cite{Schwarz}.
However, in real material systems, where the orbital moment at the edge atoms does not vanish, the mechanical effect may be absent or much smaller compared to the estimates provided by Sonin\cite{Sonin2}. The calculation of ESCs from first-principles as presented in the following, accompanied by some further developments to obtain mechanical toques from ionic forces, might allow to explore this problem at the quantitative level in future works. \\
More recently, a few works proposed the use of optical methods\cite{JWang,Werake} and,  
specifically, that spin currents can be probed by polarized light beams\cite{JWang} or via second-order nonlinear optical effect\cite{Werake}. The results refer to out-of-equilibrium spin-currents generated, for example, through laser pulses, but 
the idea should apply to ESCs as well. The possibility to exploit optical methods to address ESCs seems also rather natural considering that ESCs emerge in gyrotropic systems, which, as such, are optically active\cite{Landau}. In fact, it might be possible to reinterpret magneto-optical responses in terms of ESCs. This is an interesting direction for further research.\\
The most promising approach to measure ESCs would be by electrical means\cite{Sun}. 
Non-dissipative charge currents in currents loops are
detected by magnetic-field measurements and, similarly, one may detect ESCs
by electric-field measurements even in the absence of any spin accumulation. This is because an ESC leads to electric polarization\cite{Sonin3}. By definition the spontaneous polarization $\mathbf{P}$ of a system is calculated by 
applying an external electric field $\mathbf{E}$ and computing the derivative of the energy at $\mathbf{E}=0$,  $\mathbf{P}=\langle d\hat{H}/d\mathbf{E} \rangle\vert_{\vert \mathbf{E}\vert=0}$.
On the other hand, the components of the ESC density $j_{i}^{a}$ are given by the derivative of the energy with respect to the SU(2) vector potential as shown in Eq. (\ref{eq.DHDA}).
In the presence of SOC the change of a vector potential component $\mathcal{A}_i^a$ is related to the applied electric field via the second of Eqs. (\ref{gauge_field}). 
Thus we find the relation between the electric polarization and the current density
\begin{equation}
P_j= \bigg\langle \frac{dH}{dE_j}\bigg\rangle = \frac{e}{mc^2}\epsilon_{ija}\bigg\langle\frac{dH}{d\mathcal{A}_{i}^{a}} \bigg\rangle=\frac{e}{mc^2 } \epsilon_{ija} j_{i}^{a}.
\end{equation}
In insulators with bulk ESCs, such as InN, which is studied in Sec. (\ref{sec.InP}), one could compute from first-principles the bulk electric polarization\cite{Bernardini} with and without SOC, relate the difference to the ESC density and then compare the results for $\mathbf{P}$ to experiments. 
In metals, this argument is probably not applicable because of metallic screening. However we can expect some extra polarization near the surfaces. The extra polarization due to SOC can probably be understood in terms of the inverse spin-Hall effect. 
If we adiabatically switch on SOC in a gyrotropic material, we will produce a spin current, and because of the inverse spin Hall effect\cite{Valenzuela}, a perpendicular charge current. 
At the end of the process the transferred charge will give the extra polarization in presence of the ESC. For a bulk ESC density of the order of $10^{17}$ eV/m$^2$, which is a rather realistic value based on the results obtained in the rest of the paper, the polarization is of the order of $10^{-8}$ C/m$^2$, that is unfortunately very small. 
Nonetheless, we think that understanding the key features of ESCs and having the possibility to predict their magnitude from first-principles could help to design experimental set-ups and select the most promising materials to eventualy address their existence.
 
\section{Spin currents from DFT}\label{sec.DFTSpinCurrents}
%Having established the physical interpretation for ESCs and their existence in gyrotropic media and at surfaces, we next proceed in Sec. \ref{sec.results} to evaluate quantitatively their magnitude for a few representative example systems.  
%The calculations are carried out via Kohn-Sham (KS) DFT as described in this section. Additinal computational details are given in Appendix \ref{sec.comp_details}.\\
Having established the physical interpretation for ESCs, we now put forward a scheme to evaluate quantitatively their magnitude in material systems.  
Specifically we use Kohn-Sham (KS) DFT. The Hamiltonian in Eq. (\ref{general_H}) then becomes the single-particle KS Hamiltonian within the local spin-density approximation (LSDA).
The external potential includes the Hartree and the exchange-correlation potential, while the Zeeman field includes the exchange correlation magnetic field. 
We employ the SIESTA package\cite{Siesta} and the SMEAGOL\cite{Rocha, Rungger,book1}  
quantum transport code. They use a linear combination of atomic orbitals basis set $\{\vert \phi_n\rangle\}_{n=1, N}$, 
where each integer $n$ stands for the atom index, the principal quantum number, the angular momentum quantum number and the magnetic quantum number; $N$ is the total number of basis orbitals. 
In general the basis states are non-orthogonal and the spin-independent overlap integrals $\Omega_{nm}=\langle \phi_n | \phi_m \rangle$ 
are the elements of the overlap matrix $\Omega$.
The Hamiltonian is expanded in the basis obtaining the matrix $H$ composed of the
$2\times 2$ spin-blocks $H_{nm}=H^\mathrm{c}_{nm}\;1_2+ \mathbf{H}_{nm} \pmb{\sigma}$ for each pairs of orbitals $n$ and $m$. $H^\mathrm{c}_{nm}$ is the charge part, while $\mathbf{H}_{nm}$ is the spin part and it is a vector of matrices $(H^x_{nm},H^y_{nm},H^z_{nm})$.
Similarly, the density matrix $\rho$ and the 
so-called energy density matrix \cite{Siesta} $F=\frac{1}{2}\left[\Omega^{-1}H\rho+ \rho H  \Omega^{-1}\right]$
are composed of the blocks $\rho_{nm}=\rho^\mathrm{c}_{nm}\;1_2+\pmb{\rho}_{nm} \pmb{\sigma}$ and
$F_{nm}=F^\mathrm{c}_{nm}\;1_2+ \mathbf{F}_{nm}\pmb{\sigma}$. \\
To derive an expression for the spin current suitable for a numerical implementation, we rewrite the
spin continuity equation in terms of the spin $\mathbfcal S_n=( \mathcal S_n^x,\mathcal S_n^y, \mathcal S_n^z)$ associated to the basis orbital $n$\cite{Nikolic}. 
In particular, following refs. \cite{book1,book2} we define the components of $\mathbfcal S_n$ in terms of the symmetrized Mulliken population
\begin{equation}
\mathcal S_n^a=\frac{\hbar}{2}\frac{(\rho^a\Omega)_{nn}+(\Omega\rho^a)_{nn}} {2},\label{S_mulliken}
\end{equation}
where $a=x,y,z$.
Then, by taking their derivative with respect to time
we obtain \cite{book1, book2}
\begin{equation}
\partial_t \mathcal S_n^a=\mathcal{I}^a_{n} +\mathcal{T}^a_{n},\label{eq:dmdt}
\end{equation}
where $\mathcal{I}_{n}^a=\sum_{m=1}^{N}\mathcal{I}^a_{nm}$ and $\mathcal{T}^a_{n}=\sum_{m=1}^{N}\mathcal{T}^a_{nm}$.
$\mathbfcal{I}_{nm}=(\mathcal{I}^x_{nm},\mathcal{I}^y_{nm},\mathcal{I}^z_{nm})$ is called the spin bond current between the orbitals $n$ and $m$
and is defined as
\begin{equation}
\mathbfcal{I}_{nm}=\mathrm{Im}\left[\mathbf{H}_{nm}\rho^\mathrm{c}_{mn}+H^\mathrm{c}_{nm}\pmb{\rho}_{mn}-\Omega_{nm}\mathbf{F}_{mn}\right].\label{Ispin}
\end{equation}
$\mathbfcal{T}_{n}=(\mathcal{T}^x_{n},\mathcal{T}^y_{n},\mathcal{T}^z_{n})$ is the torque acting on $\mathbfcal{S}_n$ 
with
\begin{equation}
\mathbfcal{T}_{nm}=2\mathrm{Re}\left[\mathbf{H}_{nm}\times\pmb{\rho}_{mn}\right].
\label{eq:taufield}
\end{equation}
If $\mathcal{I}^a_{nm}$ is positive, it will describe the spin-$a$ current due to the flow into the orbital $n$ from the orbital $m$. 
In contrast, if $\mathcal{I}^a_{nm}$ is negative, it will describe the spin-$a$ current due to the flow out of the orbital $n$ towards the orbital $m$. 
It is then straightforward to verify that $\mathcal{I}^a_{nm}=-\mathcal{I}^a_{mn}$.\\
Eq. (\ref{eq:dmdt}) represents the equivalent in orbital representation of Eq. (\ref{covariant_conservation2}). We note that $\mathbfcal{T}_{n}$ in Eq. (\ref{eq:taufield}) contains the contributions from both SOC and the Zeeman (exchange-correlation) field, which were previously separated in Eq. (\ref{covariant_conservation2}).\\
Equilibrium bond currents are obtained inserting the equilibrium density matrix and energy density matrix into Eq. (\ref{Ispin}). 
Assuming the calculations to be carried out for a rectangular cuboid supercell, the global ESC per supercell is expressed as a pseudotensor
\begin{equation}
\left( \begin{array}{ccc}
  I^x_x  & I^x_y & I^x_z  \\
I^y_x & I^y_y & I^y_z\\
 I^z_x & I^z_y & I^z_z
 \end{array} \right).\label{ESCpseudo}
\end{equation}
Each component $I^a_i$ is obtained by summing the spin-$a$ bond currents $\mathcal{I}^a_{nm}$ connecting the pairs of orbitals $n$ and $m$ located on the opposite sides of the supercell surface with normal along the $i$ Cartesian direction. Details on how the calculations are practically carried out are given in Appendix \ref{sec:implementation}.
The results are then rescaled for unit cells of arbitrary shapes, or, alternatively, converted into the ESC densities $j^a_i$ defined in the previous section.
We note that bond and therefore also global spin current components have the unit of an energy. 
However, it is sometimes useful to express them in the same unit as the charge current. This can be done by multiplying the bond currents in Eq. (\ref{Ispin}) by $e/\hbar$. \\
The bond current method was first introduced within the tight-binding approach for models \cite{Todorov,Nikolic,Theodonis} and, recently, used in KS-DFT with localized basis orbitals to evaluate spin-transfer torque \cite{book1,Xie} and spin Hall effect \cite{Wang,Wesselink} in nano-devices from first-principles. 
Bond currents have however no physical meaning per se in KS-DFT calculations.
In fact, their values depend on the specific choice of the basis set and on the specific population used to define the local spin $\mathbfcal{S}_n$ [for example here we opted for the symmetrized Mulliken population of Eq. (\ref{S_mulliken})]. 
Furthermore, we have implicitly assumed that the basis set is complete, but this is never the case in practical numerical calculations. 
In spite of these issues, total currents are well defined quantities,
since the total spin $\mathbfcal{S}_\mathrm{tot}$ of a cell does not depend on the local population. 
Furthermore, the inspection of the bond currents generally provides useful physical insights into the transport properties of a system
in the same way as the local population analysis helps to understand the electronic structure. 
We will then use bond spin currents to analyse inter-atomic local current distributions.

\section{DFT results}\label{sec.results}

We employ KS-DFT with the bond current method to estimate how large ESCs are in few representative systems and to understand how the general phenomenoly described in Sec. \ref{sec.spin_currents} manifests in real materials. 
Specific computational details are given in Appendix \ref{sec.comp_details}. 
We first present results for metallic structures analysing the emergence of the (almost) universal surface ESCs and questioning whether they originate from bulk or surface states. 
We then go on studying insulating materials and demonstrating that the appearance of ESCs is by no means limited to metals as long as the system has gyrotropic symmetry. 
Finally we estimate the magnitude of ESCs in tellurium, a prototypical gyrotropic material already widely studied in the context of other gyrotropy-related effects \cite{Nomura,Vorobev,Furukawa,Tsirkin}. 

\begin{figure}
\begin{minipage}[ht]{0.19\textwidth}
\includegraphics[width=\linewidth]{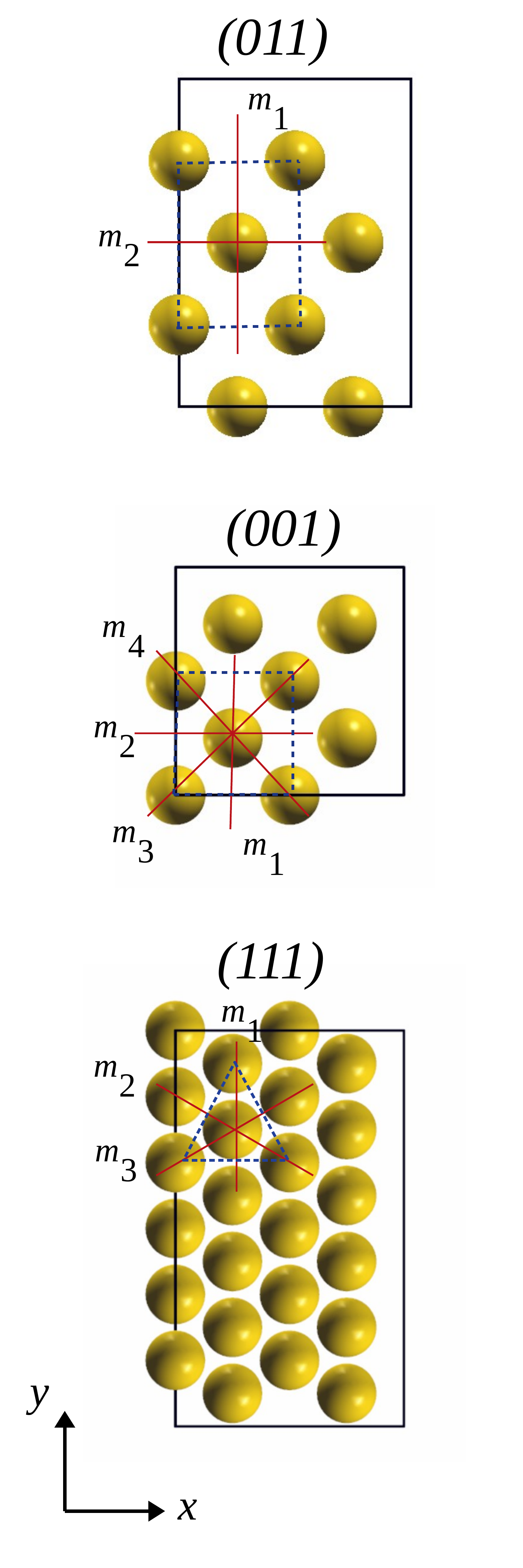}
\end{minipage}
\begin{minipage}[ht]{0.2\textwidth}
\begin{equation*}
\left( \begin{array}{ccc}
  0  & 3.3 & 0  \\
-2.55 & 0 & 0\\
 0 & 0 & 0
 \end{array} \right)
\end{equation*}
\\
\bigskip
\bigskip
\bigskip

\begin{equation*}
\left( \begin{array}{ccc}
  0  & 2.45 & 0  \\
-2.45& 0 & 0\\
 0 & 0 & 0
 \end{array} \right)
\end{equation*}
\\
\bigskip
\bigskip
\bigskip
\bigskip
\begin{equation*}
\left( \begin{array}{ccc}
  0  & 2.4 & 0  \\
-2.4 & 0 & 0\\
 0 & 0 & 0
 \end{array} \right)
\end{equation*}

\end{minipage}

\caption{Left: top view of the Au(001), Au(011) and Au(111) surfaces. To carry out the calculations we use rectangular $2 \times 2$ supercells in the surface $xy$ plane.
Right: calculated ESC pseudotensor as defined in Eq. (\ref{ESCpseudo}). 
The components are expressed in meV per surface unit cell. The surface unit cells are represented by the dashed blue lines.  
The components $I_z^x$, $I_z^y$ and $I_z^z$ are zero because $z$ is the direction normal to the surface and there can not be current flowing into the vacuum. 
Beside that, the structure of the ESC pseudotensor can be determined by analysing the surface symmetry, 
as demonstrated in appendix \ref{app.Au111} for Au(111) and in appendix \ref{App.Symmetry} for Au(011) and Au(001).
The mirror reflection lines of the surface point groups are represented as red lines (see also appendices \ref{app.Au111} and \ref{App.Symmetry}).}
\label{fig.Au}

\end{figure}

\begin{figure}[ht]
\centering\includegraphics[width=0.42\textwidth,clip=true]{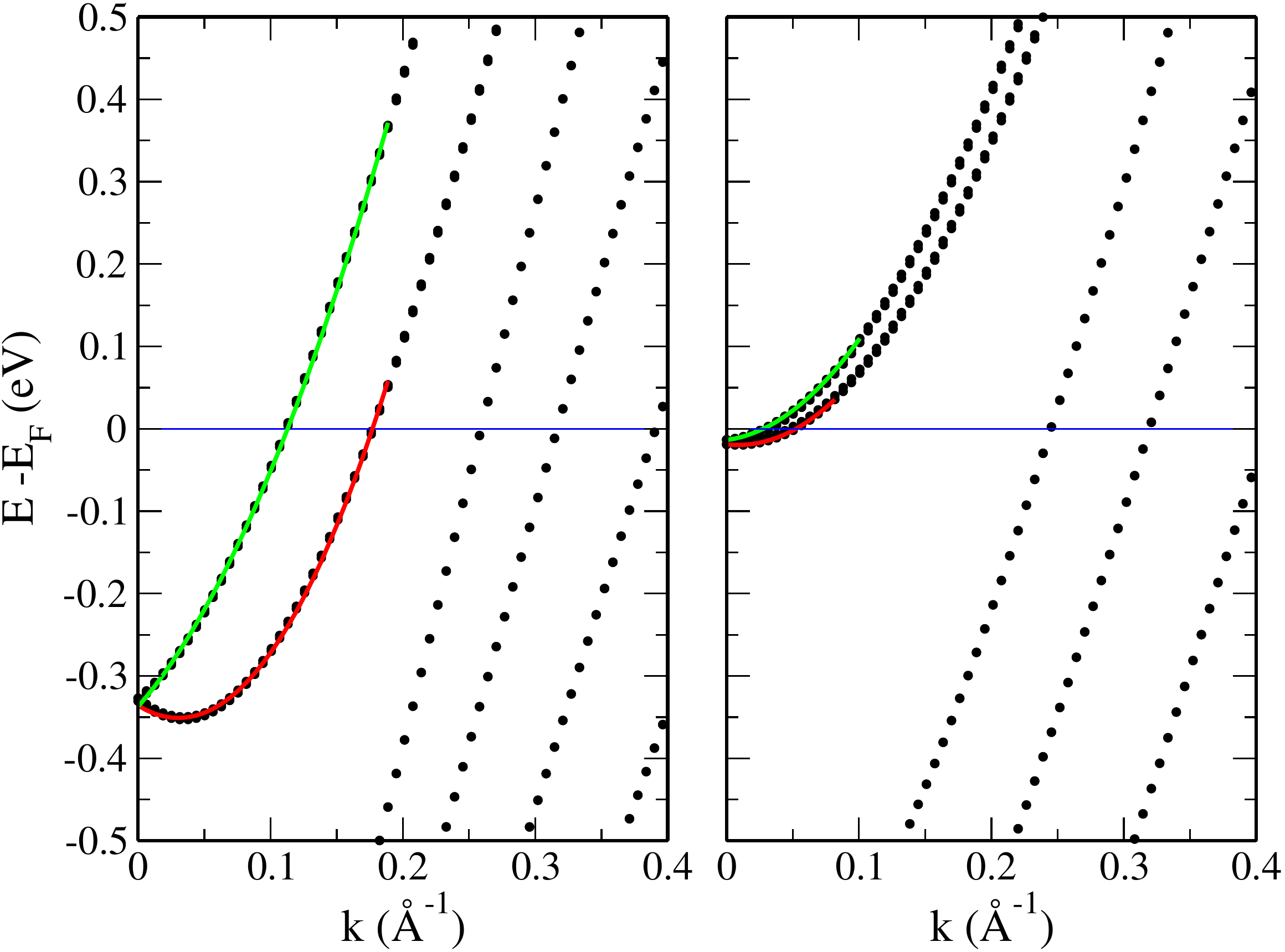}
\caption{Band structure of Au(111) (left) and Ag(111) (right). The red and green lines are the fitted Rashba bands. }
\label{fig.bands}
\end{figure}

%{table}
%{%
%\begin{tabular}{ M{1.2cm} | M{1.6cm} |M{1.6cm} |M{1.6cm} }\hline
% (100) \space  &\space   $I_{x}^{y}=-I_{y}^{x}$\space  &  0.127 eV  \space & 4.917 $\mu A$  \\\hline
% (011) \space  &\space   $I_{y}^{x}$ \space &  0.128 eV  \space &   -4.95 $\mu A$  \\
%  \space  & \space  $I_{x}^{y}$ \space &  0.165 eV  \space &  6.398 $\mu A$  \\\hline
% (111) \space  & \space  $I_{x}^{y}=-I_{y}^{x}$ \space &  0.219 eV  \space &  8.491 $\mu A$  \\\hline
%\end{tabular}}
%\caption{Non-zero components of the spin current tensor for the three Au surfaces. The units are eV, which can be converted $\mu A$ by multiplying by $e/h$. The current density in Fig. \ref{fig.Au} is obtained by dividing for the cell lenth either in the $x$ or $y$ direction. (CHECK factor $8\pi$ in the code and in the values reported here)}\label{Tab.Au}
%\end{table}

%\begin{table}
%{
%\begin{tabular}{ M{1.2cm} | M{1.6cm}|M{1.6cm} }\hline
% (100) \space  &\space   $j_{y}^{x}=-j_{y}^{x}$\space  &  $82$  A/cm  \\\hline
% (011) \space  &\space   $j_{x}^{y}$ \space &    $-61$    A/cm  \\
%  \space  & \space  $j_{y}^{x}$ \space &   $111$ A/cm  \\\hline
% (111) \space  & \space  $j_{y}^{x}=-j_{x}^{y}$ \space &  $81$ A/cm \\\hline
%\end{tabular}}
%\caption{Surface spin-current density }\label{Tab.Au}
%\end{table}

\subsection{Metallic surfaces}\label{sec.Au}

The $4d$ and $5d$ transition metals have fcc, bcc and hcp centrosymmetric crystals, and therefore bulk ESCs are absent despite the large atomic SOC. Nonetheless, ESCs emerge at surfaces and interfaces. For example, Fig. \ref{fig.Au} displays the  ESC pseudotensor of Eq. (\ref{ESCpseudo}) calculated for the three common gold surfaces Au(001), Au(110) and Au(111),
with gyrotropic point groups $\mathbb{C}_{4v}$, $\mathbb{C}_{2v}$ and $\mathbb{C}_{3v}$ (note that we use rectangular $2 \times 2$ supercells in the surface plane to carry out the calculations, but the results are presented in meV per surface unit cell). In all cases, the calculated ESCs are completely confined within the first three atomic layers underneath the surface, 
and the structure of the pseudotensor is dictated by the surface symmetry, as demonstrated in appendices \ref{app.Au111} for Au(111) and \ref{App.Symmetry} for Au(011) and Au(001). Our first-principles numerical results for real material surfaces support the general phenomenology described in Sec. \ref{sec.spin_currents}. 
Besides, we note that these surface ESCs should be subtracted from the total spin current to obtain the transport contribution when performing atomistic calculations of spin-charge conversion at surfaces and interfaces\cite{Wang} otherwise the magnitude might be overestimated.\\
We focus in particular on Au(111), which is a paradigmatic system to understand SOC-driven effects. This is because of its Shockley $L$-gap surface bands \cite{LaShell,Hoesch,Nicolay}, 
which can be mapped into the eigenenergies of the 2D Rashba model \cite{Petersen,Nicolay,Henk,Bihlmayer,Heide}. There is an ESC associated to these surface bands and it can be calculated  
by means of Eqs. (\ref{SC_xy_Rasbha_Dresselhaus_tot1}) and 
(\ref{SC_xy_Rasbha_Dresselhaus_tot2}). The results from the model can then be compared to those in Fig. \ref{fig.Au}. Since the bond current method provides the total ESC summed over {\it all} bands, bulk as well as surface bands, the proposed comparison will eventually reveal how important the contribution of the Rashba-like surface bands is
and whether an effective 2D Rashba model description is adequate to account for the main phenomenology.
As already mentioned, this is a very important open question in the wide context of SOC-driven interfacial phenomena, which have so far been described considering only spin-textured surface bands and completely neglecting bulk states \cite{Sanchez,Rousseau,XChen}.\\
The band structure of Au(111) along the $\Gamma$-$L$ direction is displayed in Fig. \ref{fig.bands} (left panel). We can clearly distinguish the Rashba-like states. 
The green and red lines are the fit to the Rashba model eigenenergies\cite{Rashba}
\begin{equation}
E_{s, \mathbf k}=\frac{\hbar^2k^2}{2m_*}+s \vert \lambda_R\vert k, \label{rashba_band}\\
\end{equation}
where $\mathbf{k}=(k_x, k_y)$ is the wave-number, $m_*$ is the electron effective mass, $\lambda_R$ the Rashba SOC constant introduced in Sec. \ref{sec.spin_currents} and $s=\pm 1$ labels the two bands, with $s=+1$ and $s=-1$ corresponding to spin-up and spin-down defined as locally perpendicular to $\mathbf k$.
The estimated $m_*$ is $0.24 m$ and $\lambda_R$ is as large as $0.99$ eV\AA. Both values are in good agreement with the results of previous  works\cite{Heide}. 
The ESC densities $j^x_{y} = -j_{x}^y=j_{R}$ associated to these Rashba bands are calculated 
using Eqs. (\ref{SC_xy_Rasbha_Dresselhaus_tot1}) and 
(\ref{SC_xy_Rasbha_Dresselhaus_tot2}) (with $\lambda_D=0$ and $m_*$ instead of $m$).
We find $j_{R}=1.25$ meV/\AA, which corresponds to Rashba surface ESC components $I^x_{\mathrm{R},y}=  -I^y_{\mathrm{R},x}= 3.6$ meV per unit cell. 
Their order of magnitude is comparable to that of the non-zero ESC components in Fig. \ref{fig.Au} (we note that any comparison, which addresses the actual numbers and not just the order of magnitude is difficult and not fully reliable because of numerical limitations). Based on our findings, we might then argue that
the surface ESC is mostly associated to the Rashba-like surface bands, and that this is a general feature of metallic surfaces. However, such conclusion is not correct.
To show that, we first extend our study to consider also silver, in particular the Ag(111) surface. \\
We assume Ag to have the same lattice constant as Au, so that, in practice, the only difference between the two systems is the atomic species.
The band structure, which is shown in Fig. \ref{fig.bands} (right panel), still presents well recognizable Rashba-like surface bands.
However, the fitted Rashba parameter $\lambda_R=0.185$ eV\AA~is rather small. Considering an estimated effective mass $m^*= 0.37 m$, this gives Rashba ESC components $I^x_{\mathrm{R},y}=  -I^y_{\mathrm{R},x}$ equal to $6\times 10^{-3}$ meV, a value three orders of magnitude smaller than in Au(111). 
These results based on the Rashba model can now be compared to the DFT calculations, which sums over all bands. 
In doing so, we find that the total component $I^x_y$ of the ESC pseudotensor is $0.6$ meV per unit cell, i.e. two orders of magnitude larger than the Rashba value $I^x_{\mathrm{R},y}$.  
Clearly, surface bands have a negligible importance in the case of Ag(111).\\
To further analyse the problem, we systematically rescale the atomic SOC by a factor $\alpha$ in our calculations. For small $\alpha$, the Rashba constant scales linearly as a function of $\alpha$. 
Then, the non-zero ESC components would scale cubically according to Eq. (\ref{SC_xy_Rasbha_Dresselhaus_tot1}). Instead we find
 that $I^x_y$ and $I^y_x$ show a linear behavior for small $\alpha$ in both Ag(111) and Au(111) (appendix \ref{ESC_alpha.111}). 
This provides an additional confirmations that bands other than the Rashba-like surface states determine surface ESCs. 
Specifically, these bands are bulk states scattered off the surface. In fact, model calculations for a semi-infinite jellium model show that the lowest order contribution to surface ESCs is due to the interference between the incident and the reflected bulk states, and is linear in the SOC constat\cite{Tokatly3}. 
This is very much reminiscent of what already observed in the case of current-induced spin-polarization, 
where the contribution of the surface states to the total surface spin-polarization is rather small, compared to the  
that of bulk states \cite{Tokatly2}. Hence, we conclude that a description of interfacial effects including only surface bands via an effective 2D Rashba model is inadequate. 
To determine ESCs in any specific situation, one must draw on detailed microscopic calculations, which take into account not just surface bands, but also the effect of the atomic SOC on bulk electronic states. \\
%, the ESC of Au(001) can not by associated at all to the Shockley surface band, which are unoccupied in this case. 
%That is a further indication that ESCs in fcc metals likely stem from bulk states although they manifest at the surface, where they are allowed by symmetry.

\begingroup
\renewcommand{\arraystretch}{1.6}
\begin{table}
{
\begin{tabular}{ M{2.2cm} | M{2.6cm}|M{2.2cm} }\hline
 Au(001) \space  &\space   $j_{\mathrm{S},y}^{x}=-j_{\mathrm{S},y}^{x}$\space  &  $20.5$  A/cm  \\\hline
 Au(011) \space  &\space   $j_{\mathrm{S},x}^{y}$ \space &    $-15.3$    A/cm  \\
  \space  & \space  $j_{\mathrm{S},y}^{x}$ \space &   $27.8$ A/cm  \\\hline
 Au(111) \space  & \space  $j_{\mathrm{S},y}^{x}=-j_{\mathrm{S},x}^{y}$ \space &  $20.3$ A/cm \\\hline
 
InP(001) $\mathbb{D}_{2d}$  \space  & \space  $j_{\mathrm{TS/BS},x}^{x}=-j_{\mathrm{TS/BS},y}^{y}$ \space &  0.425 A/cm \\\hline
InP(001) $\mathbb{D}_{2d}$  \space  & \space  $j_{\mathrm{BS},x}^{y}=-j_{\mathrm{BS},y}^{x}$ \space &  -75 A/cm \\\hline 
InP(001) $\mathbb{D}_{2d}$  \space  & \space  $j_{\mathrm{BS},z}^{y}=-j_{\mathrm{BS},x}^{z}$ \space & -57.5 A/cm \\\hline 
InP(001) $\mathbb{D}_{2d}$ \space  & \space  $j_{\mathrm{TS},x}^{y}=-j_{\mathrm{TS},y}^{x}$ \space &  75 A/cm \\\hline 
InP(001) $\mathbb{D}_{2d}$ \space  & \space  $j_{\mathrm{TS},z}^{y}=-j_{\mathrm{TS},x}^{z}$ \space & 57 A/cm \\\hline

InP(001) $\mathbb{C}_{2v}$ \space  & \space  $j_{\mathrm{BS},x}^{x}=-j_{\mathrm{BS},y}^{y}$ \space & 1.6  A/cm \\\hline  
  InP(001) $\mathbb{C}_{2v}$ \space  & \space  $j_{\mathrm{BS},x}^{y}=-j_{\mathrm{BS},y}^{x}$ \space & -74  A/cm \\\hline      
 InP(001) $\mathbb{C}_{2v}$ \space  & \space  $j_{\mathrm{TS},x}^{x}=-j_{\mathrm{TS},y}^{y}$ \space &  -1.2 A/cm \\\hline  
  InP(001) $\mathbb{C}_{2v}$ \space  & \space  $j_{\mathrm{TS},x}^{y}=-j_{\mathrm{TS},y}^{x}$ \space & 74.8 A/cm \\\hline     
    
InP(011)  \space  & \space  $j_{\mathrm{TS},x}^{x}$ \space & 55.8 A/cm \\\hline     
InP(0110)  \space  & \space  $j_{\mathrm{TS},x}^{y}$ \space & 120.3 A/cm \\\hline     
   InP(011)  \space  & \space  $j_{\mathrm{TS},y}^{x}$ \space & -94.2 A/cm \\\hline\hline    
   
 InN \space  & \space  $j_{y}^{x}=-j_{x}^{y}$ \space &  $0.37$ MA/cm$^2$
 \\\hline
 
 Te \space  & \space  $j_{y}^{x}=j_{y}^{y}$ \space &  $42.5$ MA/cm$^2$
 \\\hline
  Te \space  & \space  $j_{z}^{z}$ \space &  $28$ MA/cm$^2$
  \\\hline
 Te \space  & \space  $j_{z}^{\perp}$ \space &  $145$ MA/cm$^2$
 \\\hline
\end{tabular}}
\caption{Non zero components of either the surface ESC density (in A/cm) or of the bulk ESC density (in MA/cm$^2$) for all investigated systems.
The subscript $\mathrm{S}$, $\mathrm{TS}$ and $\mathrm{BS}$ stands for surface, top surface and bottom surface as defined in Secs. \ref{sec.Au} and \ref{sec.InP}.
}\label{Tab.CurrentDensity}
\end{table}
\endgroup

\subsection{ Zincblende and wurtzite semiconductors}\label{sec.InP}
ESCs are not unique to metals. They likewise emerge in semiconductors, as long as the SOC is not negligible and the crystal symmetry is gyrotropic. 
This is because all occupied bands, and not just those crossing the Fermi energy, can contribute to ESCs. The presence of a band gap at the Fermi energy is therefore irrelevant.
To demonstrate this, we present calculations of ESCs for semiconductors.  
In particular, we compare two of the most common crystal structures, 
namely zincblende and wurtzite. %to better grasp the role of symmetry. 
We consider InP and InN as representative examples because of the large SOC of the In atoms.\\
InP has a zincblende structure and point group $\mathbb{T}_d$, which is non-centrosymmetric, but also not gyrotropic. 
As such, there are no global bulk ESCs, although we observe that 
individual spin bond currents are non-zero (see appendix \ref{App.Bond_InP}). The situation will drastically change if we consider nanostructures.
\begin{figure}
\begin{minipage}[ht]{0.25\textwidth}
\includegraphics[width=\linewidth]{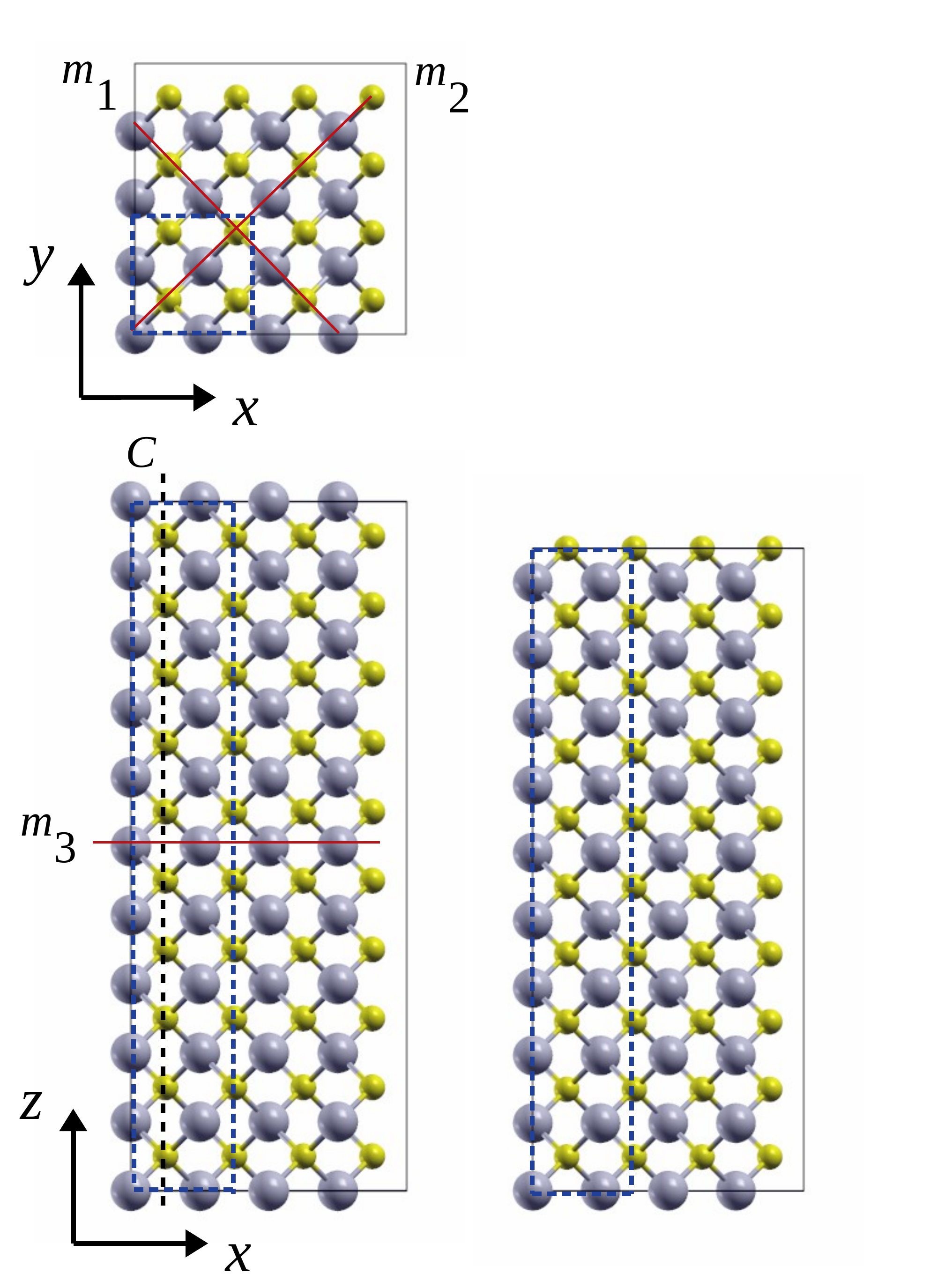}
\end{minipage}
\begin{minipage}[ht]{0.2\textwidth}
\begin{equation*}
\left( \begin{array}{ccc}
 0.22  & 0 & 0  \\
0 & -0.22 & 0\\
 0 & 0 & 0
 \end{array} \right)
\end{equation*}
%\caption*{Equal termination}
\bigskip
\bigskip
%\caption*{Different termination}
\begin{equation*}
\left( \begin{array}{ccc}
  0.1   &  -0.06 & 0  \\
0.06 &  -0.1 & 0\\
 0 & 0 & 0
 \end{array} \right) 
\end{equation*}

\end{minipage}

\caption{Left: InP(001)-oriented slabs with and without the same top and bottom terminations. The large gray spheres and small yellow spheres are respectively the In and P atoms. The calculations are carried out for $2\times 2$ supercells in the $xy$ plane. 
The surface unit cell is contained inside the blue dashed rectangle. The mirror reflection lines are in red (see appendix \ref{App.Symmetry}).
Right: Corresponding ESC pseudotensors 
as defined in Eq. (\ref{ESCpseudo}) and expressed in meV per unit cell. The components $I_z^x$, $I_z^y$ and $I_z^z$ are zero because $z$ is the normal direction to the slab surfaces.
The symmetry analysis of the ESC pseudotensors is presented in appendix \ref{App.Symmetry}.}
\label{fig.InP_100}

\end{figure}

\begin{figure}[h]
\centering\includegraphics[width=0.47\textwidth,clip=true]{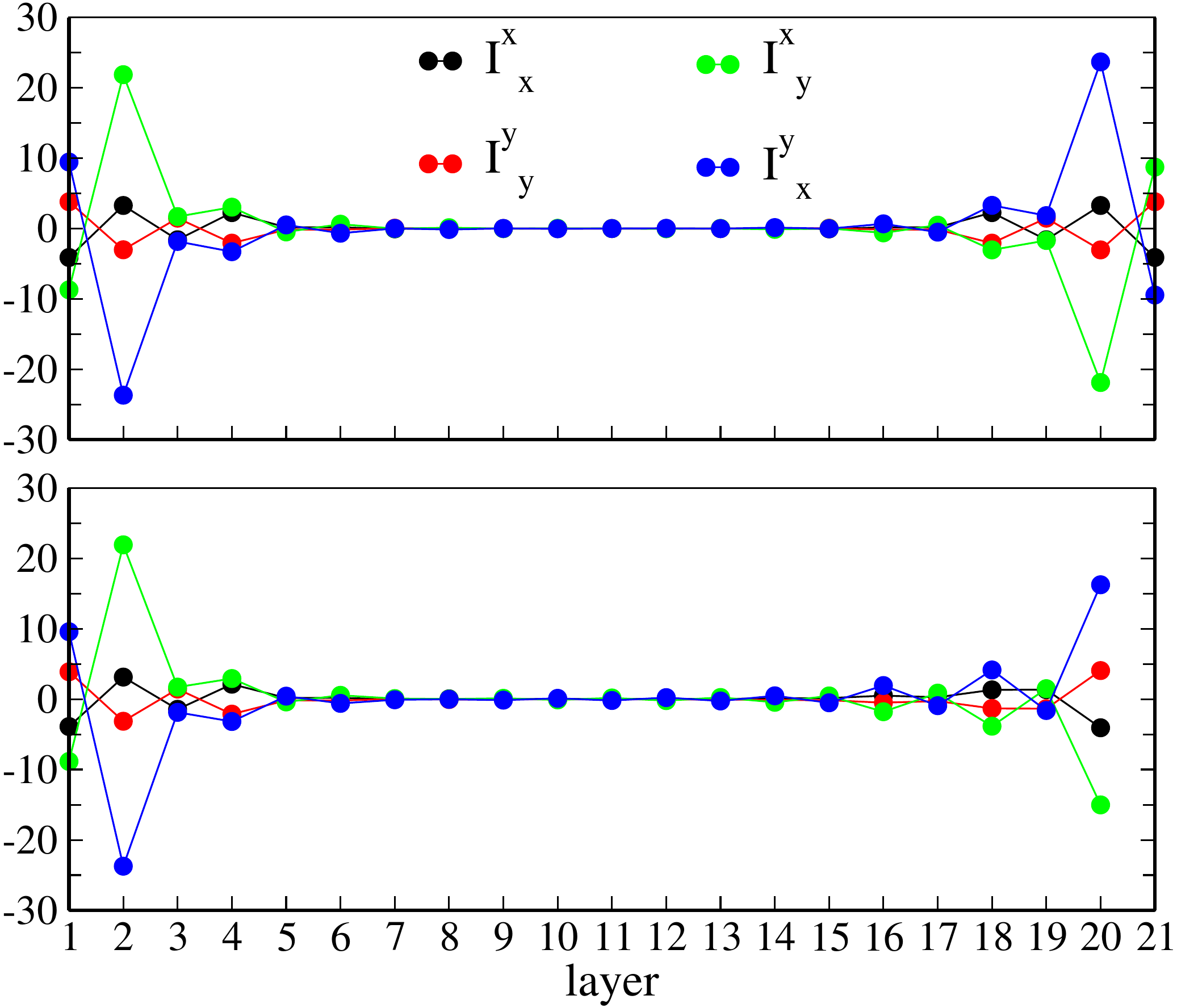}
\caption{Layer-resolved components of the ESC for the InP(001)-oriented slabs (units meV).
Top: 21-layer slab. Bottom: 20-layer slab. At each surface, the various ESC components are related by the surface symmetry operations as demonstrated in appendix \ref{App.Symmetry}.}
\label{fig.InP_100_layer}
\end{figure}

\begin{figure}
\begin{minipage}[ht]{0.18\textwidth}
\includegraphics[width=\linewidth]{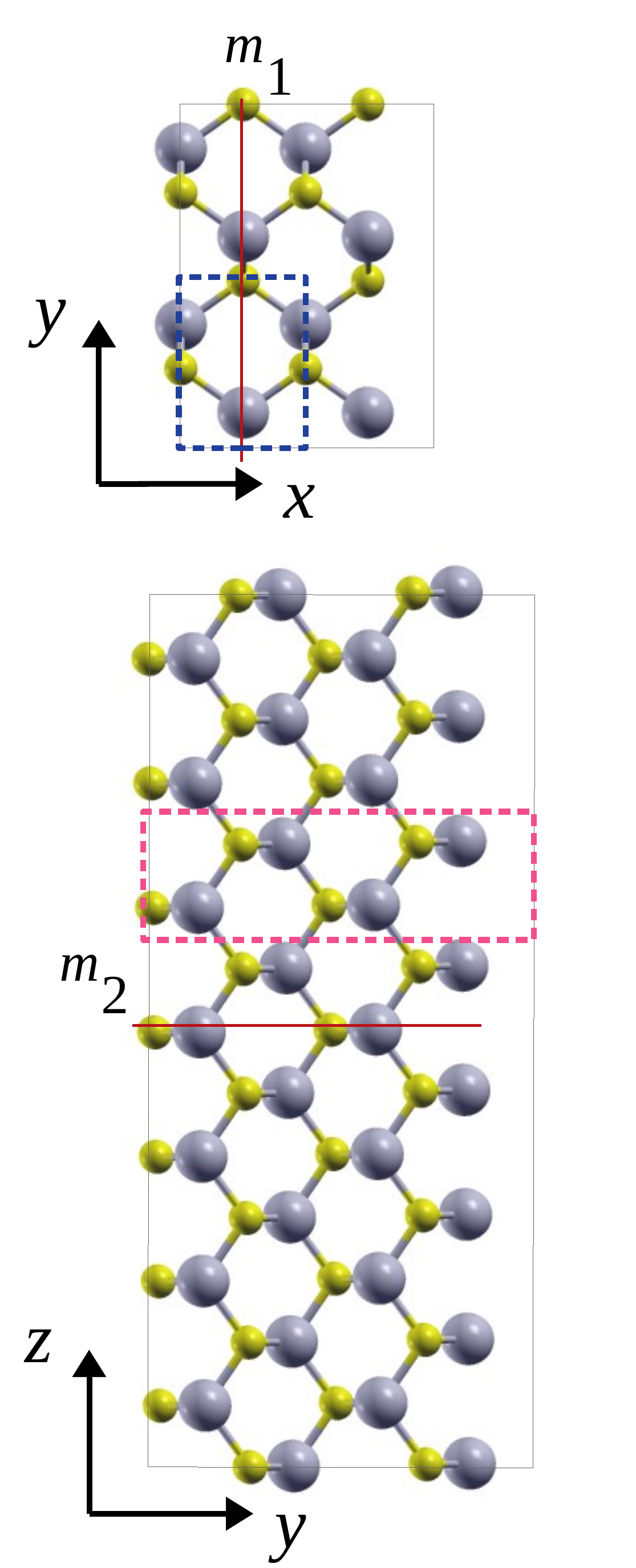}
\end{minipage}
\begin{minipage}[ht]{0.2\textwidth}
\begin{equation*}
\left( \begin{array}{ccc}
 0  & 0 &   \\
0 & 0 & 0\\
 -1.80\times 10^{-2} & 0 & 0
 \end{array} \right) 
\end{equation*}

\end{minipage}

\caption{Left: InP(110)-oriented slab. The large gray spheres and small yellow spheres are respectively the In and P atoms. 
The calculations are carried out for a $2\times 2$ supercell in the $xy$ plane. 
The surface unit cell is delimited by the blue dashed rectangle. 
The mirror reflection lines are in red (see appendix \ref{App.Symmetry}). 
The bulk InP unit cell, which is composed of two parallel atomic layers along (001), is shown inside the magenta dashed rectangle. 
Right: corresponding ESC pseudotensor as defined in Eq. (\ref{ESCpseudo}) and expressed in meV per unit cell. 
The components $I_z^x$, $I_z^y$ and $I_z^z$ are zero because $z$ is the normal direction to the slab surfaces.
The structure of the ESC pseudotensor can be further understood based on the symmetry analysis in appendix \ref{App.Symmetry}.}
\label{fig.InP_110}

\end{figure}

\begin{figure}[h]
\centering\includegraphics[width=\linewidth,clip=true]{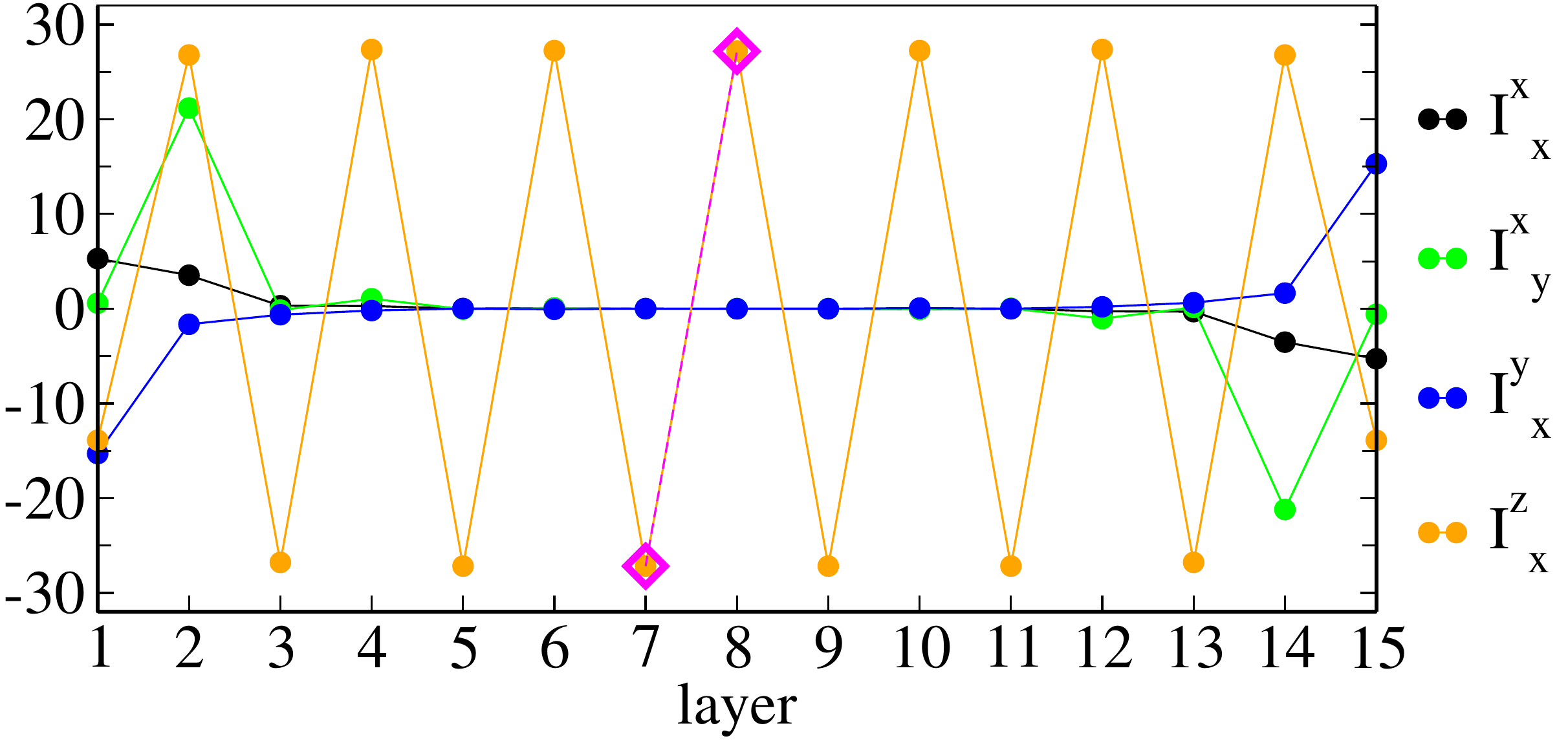}
\caption{Layer-resolved ESC for the InP(110)-oriented slab (units meV).
$I^y_y$ is zero for all atomic layers. The magenta diamonds correspond to the layer-resolved $I^z_x$ components for the bulk InP unit cell (which is shown inside the magenta dashed rectangle in the  Fig. \ref{fig.InP_110}). }
\label{fig.InP_110_layer}
\end{figure}

An InP(001)-oriented slab, which presents the same atomic termination at the two surfaces, has gyrotropic $\mathbb{D}_{2d}$ point symmetry. This
further reduces to $\mathbb{C}_{2v}$ in case of different terminations. ESCs are then allowed by symmetry in both systems. 
We consider 20-layer and 21-layer slabs. 
The calculated ESC pseudotensors of Eq. (\ref{ESCpseudo}) are reported in Fig. \ref{fig.InP_100}, together with the crystal structures of the slabs.
The non-vanishing components can be determined using general symmetry arguments. This is discussed in detail in appendix \ref{App.Symmetry}.
Here, we focus instead on a more interesting property. 
The total slab ESC is the sum of the current at the top surface (TS) and bottom surface (BS), whereas there are no ESCs flowing through the middle of the slab. 
This is seen in Fig. \ref{fig.InP_100_layer}, which displays the ESC components resolved per atomic layer.
They reach the largest absolute value either at the first or second surface layer. 
%This is similar to what we already found in the case of Au. 
%At each surface, the various ESC components are related by the surface symmetry operations as demonstrated in appendix \ref{App.Symmetry}). 
%Here we highlight instead that 
Importantly, the top-down slab symmetry in the 21-layer slab imposes that $I^{y(x)}_{x(y),\mathrm{TS}}=-I^{y(x)}_{x(y),\mathrm{BS}}$ and $I^{x(y)}_{x(y),\mathrm{TS}}=I^{x(y)}_{x(y),\mathrm{BS}}$ (see also appendix \ref{App.Symmetry} and in particular Fig. \ref{fig.InP_100_sym}). Thus, $I^{y(x)}_{x(y),\mathrm{TS}}$ and $I^{y(x)}_{x(y),\mathrm{BS}}$ cancel out, whereas $I^{x(y)}_{x(y),\mathrm{TS}}$ and $I^{x(y)}_{x(y),\mathrm{BS}}$ add up to give $I^{x(y)}_{x(y),\mathrm{slab}}=2I^{x(y)}_{x(y),\mathrm{TS}}=0.22$ meV for the whole slab. This result serves as a clear example of the distinctive feature of ESCs compared to diamagnetism. 
Diamagnetic charge currents at the bottom and top surfaces of a slab would cancel out 
satisfying the Bloch-Bohm theorem. Surface ESCs can instead contribute to give an overall finite slab ESC.\\
For a more quantitative analysis, we extract the surface ESC densities and convert them in A/cm. The results are reported in Table \ref{Tab.CurrentDensity}. 
Notably, some surface ESC densities are larger at InP(001) surfaces than at Au surfaces, 
in spite of In and Au having atomic number $49$ and $79$ respectively. 
The magnitude of ESCs can not be guessed based solely on the atomic species, but it results from the complex interplay of atomic SOC, system symmetry and electronic structure.
First-principles calculations are the only reliable way to quantitatively estimate ESCs. \\
An inspection of the electronic structure of the InP(001)-oriented slabs indicates that they are metallic with some quantum well bands crossing the Fermi energy. 
The presence of surface currents might be attributed to that. 
We therefore extend our investigation to $(011)$-oriented slabs, which are found to be semiconducting. 
In particular, we consider the slab with 15 atomic layers displayed in Fig. \ref{fig.InP_110}. 
The symmetry is $\mathbb{C}_{2v}$. Therefore we find a global ESC even though the system has no metallic bands. 
The structure of the ESC pseudotensor is again understood based on the point group operations (see appendix \ref{App.Symmetry}), but the layer-resolved analysis of the various ESC components provides more compelling physical insights. 
Fig. \ref{fig.InP_110_layer} shows that $I^x_{y}$ and $I^y_{x}$ have the largest absolute value in the atomic layers at the top and at the bottom surfaces, 
while they are negligible in the central layers. 
Furthermore $I^{y(x)}_{x(y),\mathrm{TS}}=-I^{y(x)}_{x(y),\mathrm{BS}}$ so that $I^{y(x)}_{x(y),\mathrm{slab}}=0$ owing to the slab top-down symmetry. 
This is the very same behavior already described for InP(001).  
Differently from that case though, we note that the ESC component $I^z_{x,\mathrm{slab}}$ is non-zero in InP(011). 
To understand this result we note that the unit cell 
of bulk InP consists of two atomic layers along the (110) direction
(see the magenta dashed rectangle in Fig. \ref{fig.InP_110}). 
Each of these layers have a finite $I^z_x$, but with opposite sign as shown by the magenta points in Fig. \ref{fig.InP_110_layer}. 
In infinite InP there would be a perfect compensation and the total $I^z_x$ per unit cell would vanish. In contrast, in the slab, which has an odd number of layers, 
we find an uncompensated bulk contribution to $I^z_{x,\mathrm{slab}}$ in addition to the surface one. 
Global bulk ESCs are therefore possible and they result from non-compensating local currents. 
This property is not peculiar to some nanostructures, but it is general for all gytropic systems, for example wurtize crystals.\\

\begin{figure}
\begin{minipage}[ht]{0.2\textwidth}
\includegraphics[width=\linewidth]{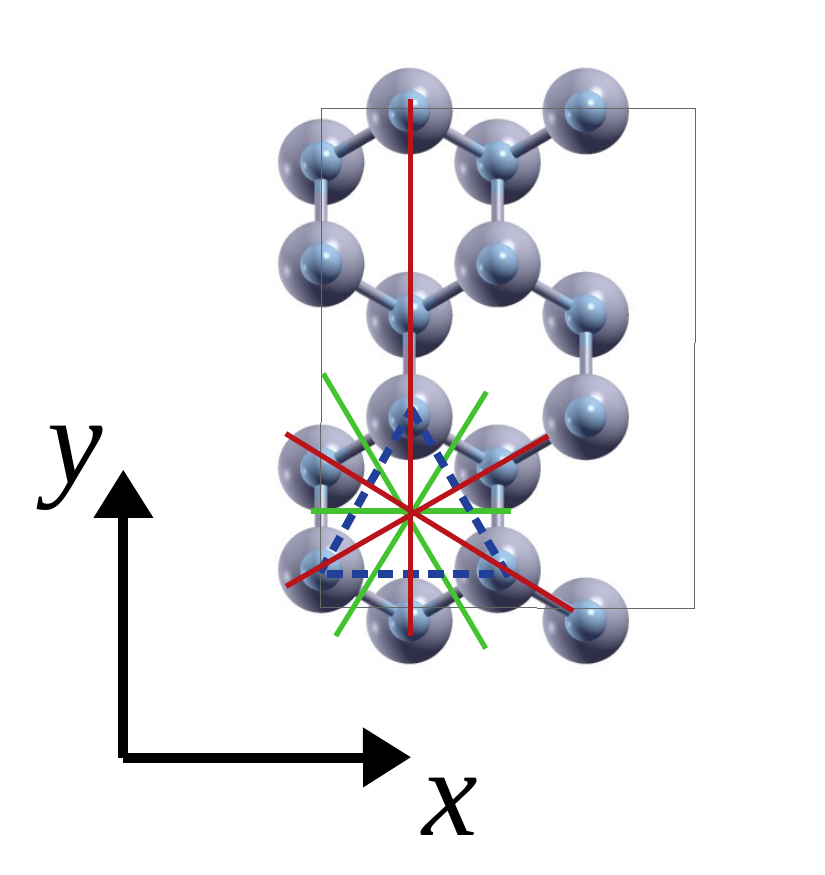}
\end{minipage}
\begin{minipage}[ht]{0.2\textwidth}
\begin{equation*}
\left( \begin{array}{ccc}
 0  & -3 & 0  \\
3 & 0 & 0\\
 0 & 0 & 0
 \end{array} \right)\times 10^{-3} 
\end{equation*}

\end{minipage}

\caption{Left: top view of the InN 2x2x2 rectangular supercell used in the calculations. 
The large and small spheres are respectively the In and N atoms. The unit cell is delimited by the dashed lines. 
The x-, y- and z-axis are respectively along the $(0001)$, the $(1\bar{1}00)$ and the $(11\bar{2}0)$ directions. There are 3 mirror planes (red lines) and 3 glide planes (green lines) in the $\mathbb{C}_{6v}$ point group. The structure of the ESC pseudotensor can be fully understood based on the corresponding symmetry operations like in the case of the Au(111) and of the InP(001) surfaces. Left: ESC pseudotensor in meV per unit cell.}
\label{fig.InN}

\end{figure}

InN is an example of wurtize semiconductor with $\mathbb{C}_{6v}$ point group. 
The ESC pseudotensor is displayed on the left hand side of Fig. \ref{fig.InN} (we note that the calculations are carried for rectangular $2\times2\times2$ supercells, but the ESC components are in meV per unit cell). 
There are two non-zero ESC components, namely $I_x^y$ and $I_x^y$, which are equal in modulus and opposite in sign as dictated by the system symmetry. 
Microscopically, the finite ESC can be understood as resulting from a non-cancellation of several bond currents. This is shown in appendix \ref{App.Bond_InN}.
Here we instead point out that $I_x^y$ and $I_x^y$ are two orders of magnitude smaller than surface ESCs in Au and InP. 
Although the wurtzite crystal structure is gyrotropic, it is obtained from the zinc-blende structure through a deformation of the tetrahedrally coordinated bond angles from a {\it cis-} to a {\it trans-} configuration. 
InP can therefore be seen as a ``mild'' gyrotropic system.
Very different results are obtained for ``strong'' gyrotropic materials, such as Te.

\begin{figure}[t!]
\centering\includegraphics[width=\linewidth,clip=true]{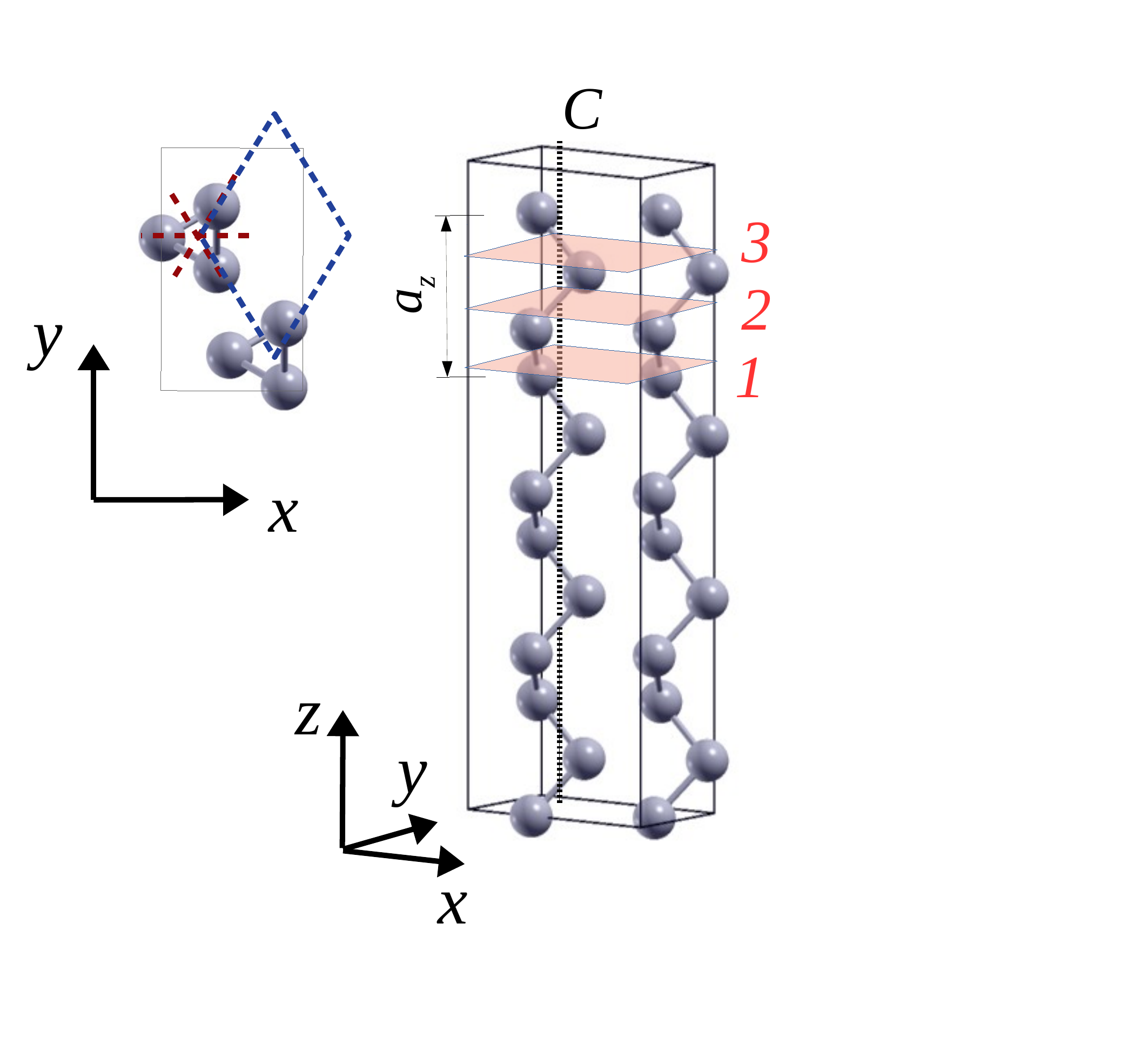}
\caption{Top view (left) and side view (right) of the rectangular supercell of right-handed Te used in the calculations. 
In the left panel the rhombohedral unit cell is delimited by the blue dashed line. 
The rotation axis $R_1$, $R_2$ and $R_3$ are represented as red dashed lines. 
The right panel also shows the three planes, which are normal to the $C$ axis (dashed black line) and through which $\mathbf I_z$ is calculated.   }
\label{fig.Te}
\end{figure}

\subsection{Tellurium}\label{sec.Te}
Te is a semiconductor.
At ambient conditions, it has a trigonal crystal structure (Te-I) consisting of weakly interacting infinite helical chains, which spiral around the $C$ axis and which can be either right- or left-handed. Each atom forms strong covalent-like intra-chain bonds with its two nearest-neighbors and weak van der Waals inter-chain bonds with its four next nearest-neighbors. The symmetry point group is $\mathbb{D}_3$, which is gyrotropic. Hence, global ESCs are allowed.\\
%The unit cell of right-handed Te contains three atoms at the positions $(0,-u,a_z)$,
%$(\sqrt{3}/2 u,1/2 u,a_z/3)$ and $(-\sqrt{3}/2 u,1/2 u, 2a_z /3)$, where $u=1.213$ \AA~is the internal atomic position parameter and $a_z=5.96$ \AA~is the lattice constant along $z$. These atoms are related by a $120$-degree rotation around $C$ followed by a translation of $(a_z/3)(0,0,1)$. This roto-translation fully determines the global spin current along $z$. In fact if we calculate $\vec I_z=(I^x_z,I^y_z,I_z^z)$ through theree consecutive non-equivalent planes $1$, $2$ and $3$ translated by $a_z/3$ along $z$ (Fig. \ref{fig.Te}) we find that it is equal to $(0,-I_z^\perp,I^z_z)$,
%$(\sqrt{3}/2 I_z^\perp,1/2 I_z^\perp,I_z^z)$ and $(-\sqrt{3}/2 I_z^\perp,1/2 I_z^\perp, I_z^z)$ with $I_z^\perp= 13$ meV and $I_z^z=3$ meV. The results are summarized in Tab. \ref{Tab.Te}.
The unit cell of right-handed Te (Fig. \ref{fig.Te}) contains three atoms at the positions $(-u,0,a_z)$,
$(1/2 u, -\sqrt{3}/2 u,a_z/3)$ and $(1/2 u, \sqrt{3}/2 u, 2a_z /3)$,
where $u=1.213$ \AA~is the internal atomic position parameter and $a_z=5.96$ \AA~is the lattice constant along $z$.
The atoms are therefore related by a $120$-degree rotation around the $C$-axis followed by a translation of $(a_z/3)(0,0,1)$. This roto-translation is the main symmetry operation that determines the global ESC along $z$, $\mathbf I_z=(I^x_z,I^y_z,I_z^z)$. 
To show that, we calculate $\mathbf I_z$ through three consecutive non-equivalent planes $1$, $2$ and $3$ translated by $a_z/3$ along $z$ (Fig. \ref{fig.Te}). The results are summarized in Tab. \ref{Tab.Te} and can be written as $(-I_z^\perp,0,I^z_z)$,
$(1/2 I_z^\perp,-\sqrt{3}/2 I_z^\perp,I_z^z)$ and $(1/2 I_z^\perp,\sqrt{3}/2 I_z^\perp, I_z^z)$, where $I_z^\perp= \sqrt{(I^x_z)^2+(I^y_z)^2}=1.05$ meV and $I_z^z=0.2$ meV. $I_z^\perp$ is the current for the spin locally parallel to the projection of the Te-Te interchain bond onto the $xy$ plane. $I_z^\perp$
is conserved along the chain, while $I_z^x$ and $I_z^y$ are not. 
This is due to a torque at every atomic site that causes $I_z^x$ and $I_z^y$ to swap along the chain as verified by using Eq. (\ref{eq:taufield}). \\
In left-handed Te the three atoms in the unit cell are at the positions $(-u,0,a_z)$,
$(-1/2 u, -\sqrt{3}/2 u,a_z/3)$ and $(-1/2 u, \sqrt{3}/2 u, 2a_z /3)$. The rotation around $C$ is of -120 degrees. Therefore $\mathbf{I}_z=(I^x_z,I^y_z,-I_z^z)$ through three planes $1$, $2$ and $3$ is respectively equal to $(-I_z^\perp,0,-I^z_z)$,
$(-1/2 I_z^\perp,-\sqrt{3}/2 I_z^\perp,-I_z^z)$ and $(-1/2 I_z^\perp,\sqrt{3}/2 I_z^\perp, -I_z^z)$.\\
Finally we note that $\mathbb{D}_3$ also contains three 90-degrees rotations $R_1$, $R_2$ and $R_3$, in addition to the roto-translation. The corresponding axis are represented by the red dashed lines in Fig. \ref{fig.Te} and determine the ESC components in the plane perpendicular to the helices. It is easy to show that $I_x^y$ and $I_y^x$ vanish, while $I_x^x$ and $I_y^y$ are finite and of identical magnitude. The calculated values per unit cell are $I^x_x=I^y_y= \pm 0.35$ meV [the $-$ ($+$) sign applies to right- (left-)handed Te]. Interestingly, these currents are of the same order of magnitude as $I_z^z$ despite the relatively large inter-chain distance $a=4.51$ \AA.\\ %Eventually they would eventually disappear if we increased $a$, while the interchain spin-currents would remain unaffected. Not True \\

\begin{table}[h]
{%
\begin{tabular}{M{1.5cm} | M{1.2cm} | M{1.2cm}|M{1.2cm} }\hline
plane & $\mathcal{I}^{x}_z$  &\space\space  $\mathcal{I}^{y}_z$ & \space\space $\mathcal{I}^{z}_z$  \\  \hline
1 & -1.05  & 0  & 0.2\\
2 & 0.53  &-0.91   & 0.2\\
3 & 0.53 &   0.91   & 0.2\\
\hline
\end{tabular}}
\caption{ESC components (in meV per unit cell) along an helical chain of Te through the three planes in Fig. \ref{fig.Te}.}\label{Tab.Te}
\end{table}

The ESC densities $j_z^\perp$, $j_z^z$, $j_x^x$ and $j_y^y$ are readily calculated and reported in Tab. \ref{Tab.CurrentDensity}. 
$j_z^\perp$ is the largest and it is of the order of $10^{-8}$ A/\AA$^2$. 
To set a reference for the magnitude, we calculate the transport spin-current through an ideal Fe/MgO/Fe magnetic tunnel junction with 4 MgO layers and with the magnetizations of the two Fe electrodes set in the parallel configuration. 
Notably, an applied bias voltage as large as $4.5$ V is required to drive a spin current density of $\sim  10^{-8}$ A/\AA$^2$ through such device\cite{book1}, which behaves as an almost perfect spin -filter \cite{Butler}. 
Hence, the case of Te indicates that bulk ESCs are by no means small compared to transport spin-currents  used in spintronics.  
Nonetheless, we remind that ESCs are not transport currents and they can not be used to read and write information.

%$J^z_z= -11.027204045 \times 10^-9$ A$/$\AA$^2$ and $J^\perp_z=5.8 \times 10^-8$ A$/$\AA$^2$

%$J^x_x=-1.672439116\times10^-8$ A$/$\AA$^2$
%$J^y_y=-17.160590829\times10^-9$ A$/$\AA$^2$

\section{Conclusions}\label{sec.conclusions}
Global ESCs are allowed by symmetry and therefore exist in the very broad class of gyrotropic materials. 
ESCs emerge in metals and insulators alike, they do not transport spin and they do not result in spin accumulation. Nonetheless ESCs should be carefully subtracted when calculating transport spin currents.
The physical origin can be uncovered by making an analogy between electronic systems with SOC and non-Abelian gauge theories.
ESCs can be identified with diamagnetic color currents appearing as the response to an effective (non-Abelian) magnetic field. 
They lead to the appearance of an electric polarization, which, although very small, could be eventually measured. \\
Systems, which are not gyrotropic, can become gyrotropic by lowering their symmetry, for example through some 
crystal deformations. ESCs are therefore quite common properties of bulk materials. Moreover they are universal at surfaces and interfaces. \\
An ESC is mathematically described in terms of a second-rank pseudotensor. Its structure in a given system is completely dictated by the system symmetry.
However, the magnitude of the components depends on the subtle interplay of atomic SOC and electronic structure.
It can not be predicted based on effective models, but only by means of accurate first-principles studies. We therefore
used DFT to compute ESCs via the bond currents method.
Calculations performed for a wide range of systems, including metallic surfaces as well as common semiconductors, 
showed that global ESCs can be quite large. In particular, in the prototypical gyrotropic material Te, we found that their magnitude is comparable to transport currents used in common spintronics applications.

\section{Acknowledgments}
A.D. and I.R. thank Maria Stamenova for useful discussions about the definition of bond currents. 
A.D. was funded through the EU Marie Sklodowska-Curie individual fellowship SPINMAN (ID SEP-210189940) during the very initial stage this work. 
The development of the work was then supported by the Science Foundation Ireland (SFI) Royal Society University Research Fellowship URF-R1-191769 and by the European Commission H2020-EU.1.2.1 FET-Open project INTERFAST (ID 965046). 
I.V.T. acknowledges support by Grupos Consolidados UPV/EHU del Gobierno Vasco (Grant No. IT1249-19) and the Spanish MICINN Project No. PID2020-112811GB-I00.
I.R. acknowledges the support of the U.K. Department of Business, Energy and Industrial Strategy (BEIS). A.R. was supported by the European Research
Council (ERC-2015-AdG694097), the Cluster of Excellence ``CUI: Advanced Imaging of Matter'' of the Deutsche Forschungsgemeinschaft (DFG) - EXC 2056 (project ID 390715994), SFB925 ``Light induced dynamics and control of correlated quantum systems'' and partially by the Federal Ministry of Education and
Research Grant RouTe-13N14839. The Flatiron Institute is a division of the Simons Foundation.

\appendix

%
%\begin{figure}[t!]
%\centering\includegraphics[width=0.38\textwidth,clip=true]{Fig_lattice_scheme2_new}
%\caption{Procedure to calculate the bond currents for a periodic system. The model simulation cell contains two atoms, $1$ and $2$. 
%We perform an inverse Fourier transform of the Hamiltonian matrix, of the the overlap matrix and of the (energy) density matrix to a real space representation.
%Once this is done, the bond currents connecting atoms inside and outside the simulation cell can be easily calculated. Here for example we draw the bond current between the atom $1$ and $2'$.}
%\label{fig.scheme2}
%\end{figure}

\begin{figure}[t!]
\centering\includegraphics[width=0.32\textwidth,clip=true]{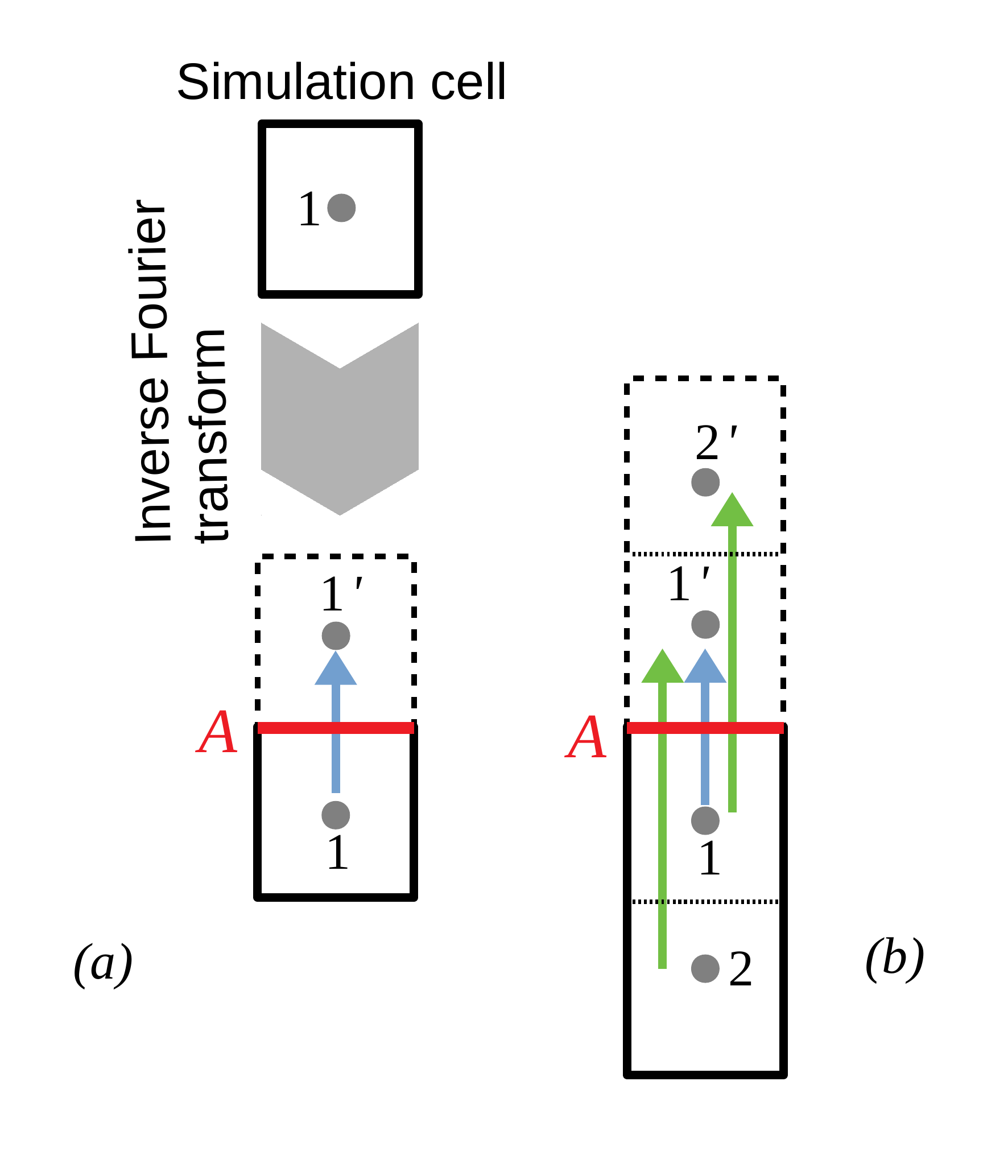}
\caption{Procedure to calculate the bond currents for a periodic system. The model simulation cell contains the atom $1$ inside a square unit cell. 
We perform an inverse Fourier transform of the Hamiltonian matrix, of the the overlap matrix and of the (energy) density matrix to a real space representation.
Once this is done, the $a$-spin component bond current $\mathcal{I}^a_{1'1}$ between $1$ and its equivalent atom $1'$ can be easily calculated. In (a) the global $a$-spin current through the surface $A$ is equal to $\mathcal{I}^a_{1'1}$. In (b) we assume that non-zero bond currents extend to second nearest neighbor atoms. Therefore we need to use the $2\times 1$ supercell in the calculation. 
The global current is given by the sum of $\mathcal{I}^a_{1'1}$, $\mathcal{I}^\alpha_{1'2}$ and $\mathcal{I}^a_{2'1}$.}
\label{fig.scheme2}
\end{figure}

\section{Implementation details for the bond currents calculations}\label{sec:implementation}
The calculation of bond currents is straightforward for a finite system. 
They are obtained through Eq. (\ref{Ispin}) inserting the Hamiltonian matrix, the density matrix and the energy density matrix. 
In contrast, some care is needed for infinite systems, such as crystals and surfaces. \\
Crystals are treated in KS-DFT by applying periodic boundary conditions. The Hamiltonian and the overlap matrices $H_{\mathbf{k}}$ and $\Omega_{\mathbf{k}}$ depend on the the wave-number $\mathbf{k}$ 
inside the Brillouin zone ($BZ$), and the eigenstates of the Schr\"odinger equation are Bloch states. The density matrix $\rho_{\mathbf{k}}$ and the energy density matrix $F_{\mathbf{k}}$ introduced in Sec. \ref{sec.DFTSpinCurrents} also depend on $\mathbf k$, and have elements $\rho_{\mathbf k,nm}$ and $F_{\mathbf k,nm}$.
Both the indices $n$ and $m$ refer to orbitals inside the simulation cell and centered at the coordinates $\mathbf{R}_n=(R_{x,n},R_{y,n},R_{z,n})$ and $\mathbf{R}_m=(R_{x,m},R_{y,m},R_{z,m})$. 
In order to obtain the current flowing in and out of that cell we have to up-fold the Hamiltonian matrix, the overlap matrix, the density matrix and the energy density matrix to real space by performing the inverse Fourier transform \cite{Siesta}
\begin{eqnarray}
  H_{n'm} =\frac{1}{N_{k}}\sum_{\mathbf{k}\in BZ}e^{-i\mathbf{k}(\mathbf{R}_{n'}-\mathbf{R}_n)}H_{\mathbf{k},nm},\\
  \Omega_{n'm} =\frac{1}{N_{k}}\sum_{\mathbf{k}\in BZ}e^{-i\mathbf{k}(\mathbf{R}_{n'}-\mathbf{R}_n)}\Omega_{\mathbf{k},nm},\\
   \rho_{n'm} =\frac{1}{N_{k}}\sum_{\mathbf{k}\in BZ}e^{-i\mathbf{k}(\mathbf{R}_{n'}-\mathbf{R}_n)}\rho_{\mathbf{k},nm},\\
  F_{n'm} =\frac{1}{N_k}\sum_{\mathbf{k}\in BZ}e^{-i\mathbf{k}(\mathbf{R}_{n'}-\mathbf{R}_n)}F_{\mathbf{k},nm}.
\end{eqnarray}
where $N_k$ is the number of k-points. $n'$ refers to the orbital equivalent to $n$, and which is centered at the coordinate $\mathbf{R}_{n'}=(R_{x,n'},R_{y,n'},R_{z,n'})$ outside the cell and related to $\mathbf{R}_{n}$ by a lattice vector translation.
The bond current connecting any two orbitals $m$ and $n'$ can then be computed by using Eq. (\ref{Ispin}). The global spin current $I^a_i$ for the spin component $a (=x,y,z)$ through the cell surface along the normal direction $i(=x,y,z)$ is obtained as 
\begin{equation}
 I^a_i=\sum_{n'> m}\mathcal{I}^a_{n'm}\,,\,\,\, \mathrm{for}\, R_{i, n'}>R_{i, m}. \label{global_current}
\end{equation} 
This procedure is illustrated in Fig. \ref{fig.scheme2}-a for a 2D model system with one atom $1$ inside a square unit cell.  
After performing the inverse Fourier transform from reciprocal to real space we obtain the spin-$a$ bond current $\mathcal{I}^a_{1'1}$ between the atom $1$ and its equivalent atom $1'$ in a neighbor cell. 
The global bond spin-current for the spin component $a$ through the cell surface $A$ is then equal to 
$\mathcal{I}^a_{1'1}$.\\
In first-principles calculations, non-zero bond currents extend generally beyond nearest neighbor atoms. Therefore, the considered cells have to be large enough to contain all orbitals $n$ and $m$ with a finite
$\mathcal{I}^a_{n'm}$. The size of the  supercell is set by the extension of the basis orbitals. This is shown in Fig. \ref{fig.scheme2}-b, where we now assume that non-zero bond currents extend to second nearest neighbor atoms. 
The global current through the surface $A$ is equal to
$\mathcal{I}^a_{1'1}$ in a calculation that only considers the unit cell. In contrast, the global current is given by the sum of $\mathcal{I}^a_{1'1}$, $\mathcal{I}^a_{1'2}$ and $\mathcal{I}^a_{2'1}$ when we properly consider a supercell with two atoms $1$ and $2$. Using the unit cell instead of the supercell would result in an error in the calculation of the global current.\\
Surfaces are studied by using the implementation of DFT based on the Green's function method \cite{Rocha,book2}. 
This allows for an effective description of systems, which 
are semi-infinite in the direction perpendicular to the surface, while periodic boundary conditions are applied only in the parallel directions. 
In practice the implementation relies on the partition of the system into the surface region (SR), with $N_{SR}$ orbitals, and the bulk region. 
The effect of the bulk on the SR is described through the embedding self-energy
$\Sigma$.
The retarded Green's function is then defined as
\begin{equation}
G_{SR}(E,\mathbf{k}_\parallel)=[(E+i\delta)\Omega_{SR,\mathbf{k}_\parallel}-H_{SR,\mathbf{k}_\parallel}-\Sigma(E,\mathbf{k}_\parallel)]^{-1}.
\end{equation}
where $\delta\rightarrow 0^+$. $\mathbf{k}_\parallel$ is the momentum parallel to the surface, $H_{SR,\mathbf{k}_\parallel}$ is the SR Hamiltonian and $\Omega_{SR,\mathbf{k}_\parallel}$ is the SR overlap matrix. 
The density matrix and the energy density matrix read
\begin{eqnarray}
 \rho_{SR,\mathbf{k}_\parallel}=\frac{1}{2\pi}\int dE f(E) A_{SR}(E,\mathbf{k}_\parallel),\\
  F_{SR,\mathbf{k}_\parallel}=\frac{1}{2\pi}\int dE E f(E) A_{SR}(E,\mathbf{k}_\parallel),
\end{eqnarray}
where
\begin{equation}
 A_{SR}(E,\mathbf{k}_\parallel)=i[G_{SR}(E,\mathbf{k}_\parallel)-G_{SR}^\dagger(E,\mathbf{k}_\parallel)]
\end{equation}
is the spectral function. 
Bond currents are evaluated after 
 up-folding  $\rho_{SR,\mathbf{k}_\parallel}$, $F_{SR,\mathbf{k}_\parallel}$, $H_{SR,\mathbf{k}_\parallel}$ and $\Omega_{SR,\mathbf{k}_\parallel}$ to real space as explained above.\\

 \section{Computational details}\label{sec.comp_details}
Our calculations are carried out with a development version of the SIESTA package\cite{Siesta} and of the SMEAGOL quantum transport code \cite{Rocha, Rungger, book1}, which is based on SIESTA\cite{Siesta}. 
We use the LSDA exchange-correlation density functional for all systems, except for Te.
Since the LSDA predicts Te to be a metal instead of a semiconductor, we use the Perdew-Burke-Ernzerhof (PBE) generalized gradient approximation (GGA) \cite{PBE1, PBE2} for this material. The SOC is included by means of the on-site approximation of Ref. \cite{Fernandez}. 
We treat core electrons with norm-conserving Troullier-Martin pseudopotentials. 
Although ESCs stem from all occupied states, we expect that core states will contribute marginally as 
they are localized very close to the nuclei. The error introduced by not including core states should be negligible.
The valence states are expanded through a numerical atomic orbital basis set including multiple-$\zeta$ and polarized functions \cite{Siesta}, which are set to zero beyond a certain cutoff radius. 
For all materials, these functions are optimized in order to closely reproduce the occupied KS band structure calculated with the Quantum Espresso plane-wave code \cite{Espresso}. We have shown in several previous works that our development versions of SIESTA and SMEAGOL are able to accurately describe materials with large SOC \cite{bise,Jakobs,Narayan}.
%We assume the KS band structure to provide a good approximation of quasi-particle band structure. 
The DFT band gap problem for semiconductors is not expected to impact our results as only occupied states contribute to ESCs. The LSDA and the GGA valence bands of InN and Te are quite well described when compared to the results obtained either by many-body perturbation theory within the GW approximation\cite{Svane,Hirayama} or by using hybrid functionals \cite{Tsirkin}. We note that there are only some slight differences in the effective masses. Addressing the impact of these differences on quantitative results is beyond the goal of this paper and it is left for possible future studies.  \\
The numerical precision of the computed bond currents depends on the $\mathbf{k}$-point grid and on the convergence threshold for the density matrix. 
We set that threshold to $10^{-6}$, which is extremely tight for SIESTA and SMEAGOL. We then systematically converge the $\mathbf{k}$-point grid until the change 
in each bond current is smaller than  $10^{-6}$ eV. Going beyond this limit is too computationally demanding. 
Furthermore, and more importantly, the computed global currents are generally several orders of magnitude larger that $10^{-6}$ eV and no better precision is therefore needed.\\
For all materials we use the experimental lattice constants unless stated otherwise, and the atomic positions are not optimized in order to prevent small reductions of the ideal systems' symmetry.
We consider rectangular supercells to simplify the evaluation of Eq. (\ref{global_current}), and we rescale the results to the values per {\it unit cell} at the end of the computation. The size of the used supercell varies from system to system. 
The lattice vectors have to be chosen larger than the cutoff radius of the basis set orbitals, as explained in detail in Appendix \ref{sec:implementation}.

\section{Transformations of the ESC components in Au(111)}\label{app.Au111}
Au(111) has $\mathbb{C}_{3v}$ point group, which contains 1) one mirror reflection through the mirror line $m_1$ parallel to the $y$-axis, 2) one reflection through the line $m_2$
forming an angle of 150 degrees with the $x$-axis, and 3) one reflection through the line $m_3$
forming an angle of 30 degrees with the $x$-axis (see the bottom left panel of Fig. \ref{fig.Au}). The structure of the ESC pseudotensor in the bottom right panel of Fig. \ref{fig.Au} can be fully understood based on these transformations, as we now outline. \\
Under the reflection through $m_1$, the components of the ESC pseudotensor transform as
\begin{equation}
{I'}^a_i=\mathrm{det}M \sum_{bj} M^a_b I^b_j M^j_i,\label{tensor_trans}
\end{equation}
where the reflection matrix $M$ is
 \begin{equation}
M=\left( \begin{array}{cc}
  -1  &  0   \\
 0  &  1 \\
 \end{array} \right).
\end{equation}
We then obtain ${I'}_x^x=-I_x^x$, ${I'}_y^y=-I_y^y$,
${I'}_x^y=I_x^y$ and ${I'}_y^x=I_y^x$. The ESC pseudotensor therefore remains invariant only if $I_x^x=I_y^y=0$.\\
Next, we consider the reflection through the mirror plane $m_3$. This is expressed through the matrix  
 \begin{equation}
M_3=\left( \begin{array}{cc}
  -1/2   &  \sqrt{3}/2   \\
 \sqrt{3}/2  &  1/2 \\
 \end{array} \right) .
\end{equation}
For this case we find that
\begin{eqnarray}
 {I'}_x^x=-\frac{1}{4} I^x_x-\frac{\sqrt{3}}{4}I^y_x-\frac{\sqrt{3}}{4}I^x_y
-\frac{3}{4} I^y_y,\\
 {I'}_y^x=-\frac{\sqrt{3}}{4}I^x_x-\frac{3}{4} I^y_x +\frac{1}{4} I^x_y+\frac{\sqrt{3}}{4}I^y_y,\\
  {I'}^y_x=-\frac{\sqrt{3}}{4}I^x_x-\frac{3}{4} I_y^x +\frac{1}{4} I^y_x+\frac{\sqrt{3}}{4}I^y_y,\\
  {I'}^y_y=-\frac{3}{4}I^x_x+\frac{\sqrt{3}}{4} I^y_x +\frac{\sqrt{3}}{4} I^x_y-\frac{1}{4}I^y_y.
 \end{eqnarray}
The ESC pseudotensor will be invariant only if $I_x^x=I^y_y=0$ and $I_x^y=-I^x_y$. A similar reasoning also applies to the reflection through $m_2$. Hence, we clearly see how the $\mathbb{C}_{3v}$ symmetry dictates the structure of the ESC pseudotensor in Fig. \ref{fig.Au}.

\begin{figure}[t!]\centering\includegraphics[width=0.4\textwidth,clip=true]{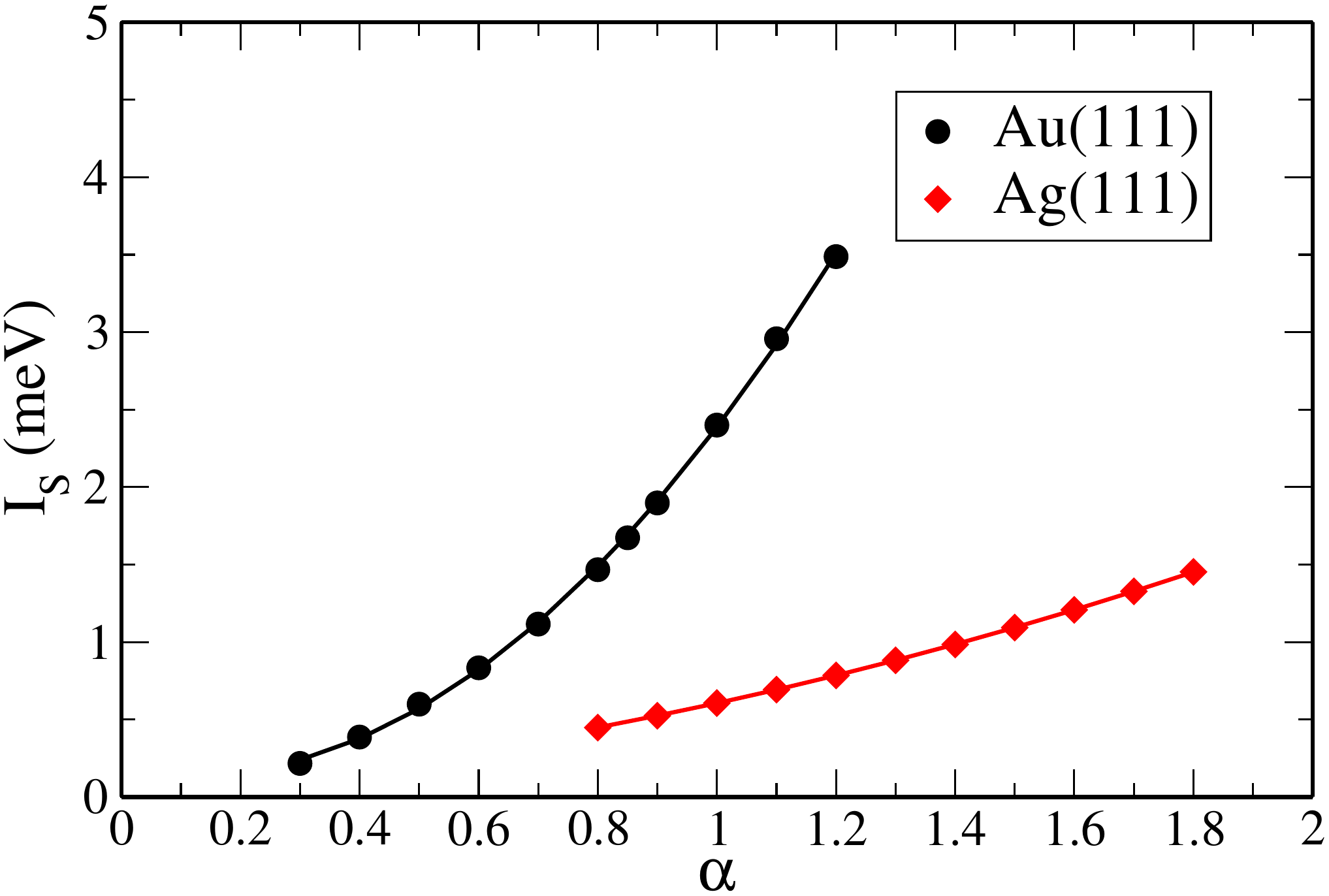}
\caption{ESC $I_S\equiv I^x_{y}=-I^y_x$ as a function of the SOC rescaling parameter $\alpha$ for the Au(111) and the Ag(111) surfaces.}
\label{fig.ESP_vs_alpha}
\end{figure}

\section{ESC components as a function of the SOC strength in Au(111) and Ag(111)}\label{ESC_alpha.111}
The SOC Hamiltonian matrix elements $V^{soc}_{nm}=\langle \phi_n \vert \hat{V}^{soc}\vert \phi_m\rangle$ 
in the DFT calculations can be re-scaled by a constant $\alpha$, that is $V^{soc,\alpha}_{nm}=\alpha V^{soc}_{nm}$. 
The surface ESC components $I_S\equiv I^x_{y}=-I^y_x$ defined in Sec. \ref{sec.Au} can then be calculated as a function of $\alpha$. 
The results are represented in Fig. \ref{fig.ESP_vs_alpha}. The data can be approximated with very high accuracy to a quadratic function $I_{S}=a_2 \alpha^2+a_1\alpha+a_0$. The fitted parameters are 
 $a_0=0.165$ meV, $a_1=-0.6$ meV, $a_2=2.83$ meV for Au(111) and $a_0=0.04$ meV, $a_1=0.29$ meV, $a_2=0.275$ meV for Ag(111).

\begin{figure}[t!]
\centering\includegraphics[width=0.35\textwidth,clip=true]{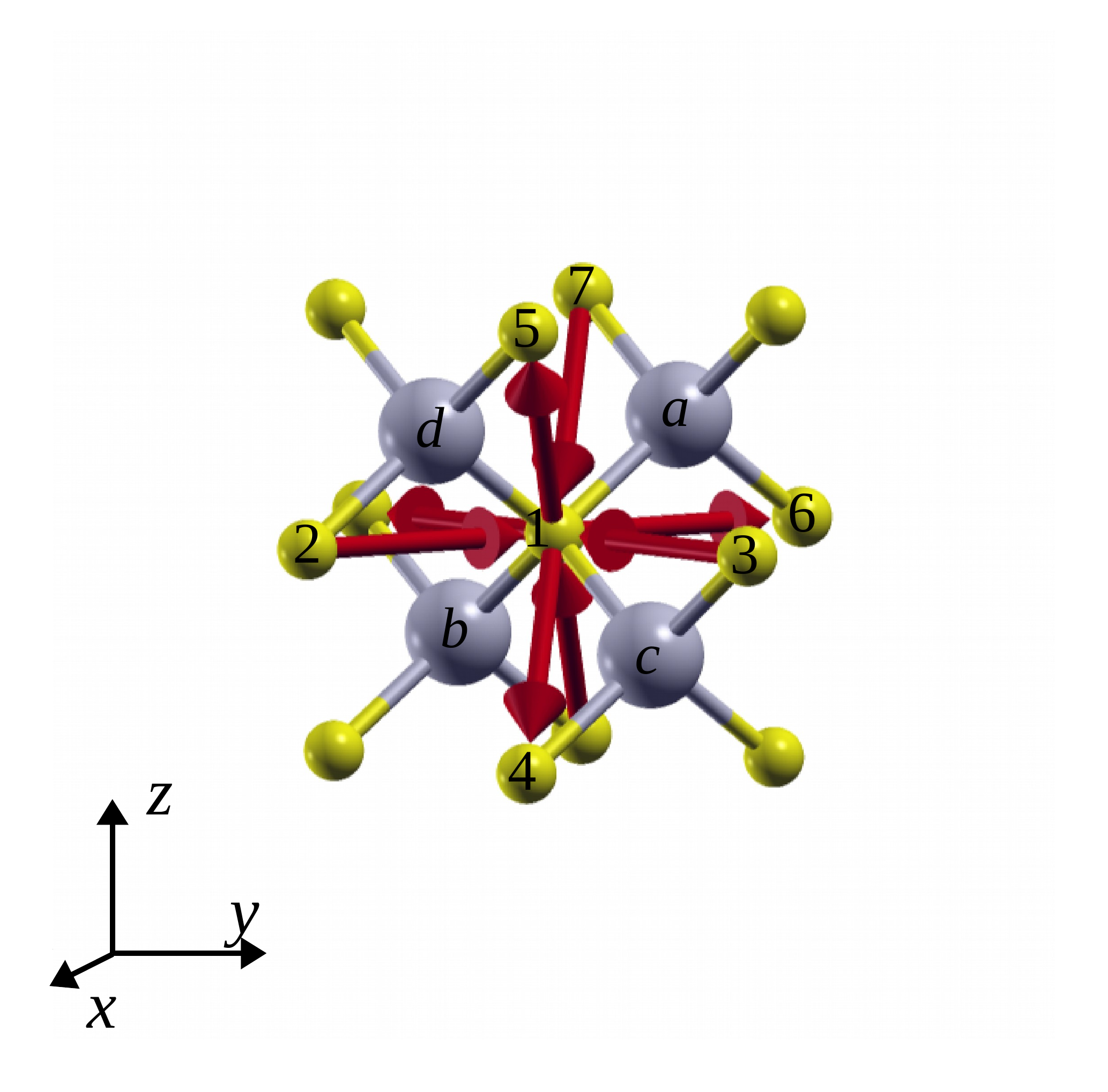}
\caption{Spin-$x$ bond currents from a P atom (labelled P$_1$) in bulk InP to its surround n.n. P atoms. The large gray spheres and small yellow spheres are respectively the In and P atoms. An arrow entering (leaving) P$_1$ means that the bond current is negative (positive). The bond currents have all the same modulus.}
\label{Fig_InP_bond_currents}
\end{figure}

\section{Bond currents in bulk InP}\label{App.Bond_InP}
InP has $\mathbb T_d$ point group.
As a result, there are no global bulk ESCs as discussed at the beginning of Sec. \ref{sec.InP}. 
Individual spin bond currents are nonetheless non-zero.
In particular, we find that the largest spin bond currents are between nearest neighbors (n.n.) P atoms despite the much larger SOC of the In atoms. These P-P n.n. bond currents are equal to $32$ meV. \\
In Fig. \ref{Fig_InP_bond_currents} we show the bond currents for the spin $x$ component, which connect an atom P$_1$ to all its n.n. P atoms 
[note that for simplicity we use a square unit cell with the $x$-, $y$- and $z$-axis along the (100), (010) and (001) directions]. 
The bond currents respect the $\mathbb T_d$ symmetry of the unit cell and they transform as pseudovectors. 
In particular, we find $\mathcal{I}^x_{12}=-\mathcal{I}^x_{14}$ and $\mathcal{I}^x_{15}=-\mathcal{I}^x_{13}$ because of the reflection through the plane bisecting In$_a$-P$_1$-In$_b$. 
At the same time we see that $\mathcal{I}^x_{15}=-\mathcal{I}^x_{12}$ because of the reflection through the In$_d$-P$_1$-In$_c$ plane.
These symmetries therefore imply that $\mathcal{I}^x_{14}$ and $\mathcal{I}^x_{15}$ respectively cancel $\mathcal{I}^x_{12}$ and $\mathcal{I}^x_{13}$ along $x$. 
The other smaller (in modulus) bond currents, which connect P$_1$ to the In atoms and to the farther P atoms, undergo identical compensations. 
Hence, the global ESC component $I_x^x$ vanishes. Similarly, $I^x_y$ and $I^x_z$ vanish as well.
This is because $\mathcal{I}^x_{13}=-\mathcal{I}^x_{16}$ under the reflection through the In$_a$-P$_1$-In$_c$ plane and $\mathcal{I}^x_{15}=-\mathcal{I}^x_{17}$ under the reflection through In$_a$-P$_1$-In$_d$ plane. 
Therefore global ESCs for the spin $x$ component do not exists along any Cartesian direction. 
The very same reasoning can be easily applied to the spin $y$ and $z$ components finally demonstrating that the ESC pseudotensor vanishes. \\

\begin{figure}[t!]
\centering\includegraphics[width=0.34\textwidth,clip=true]{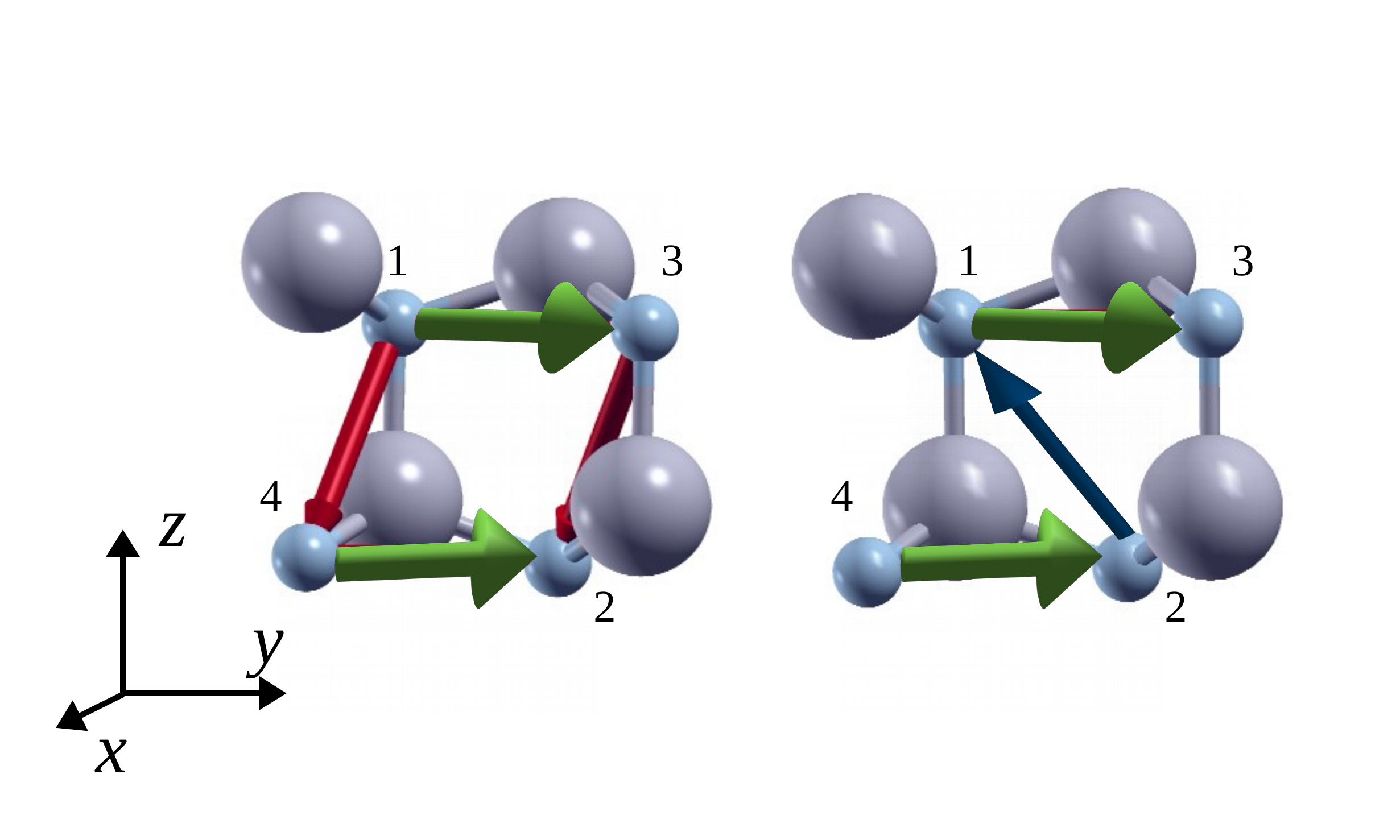}
\caption{Spin-$x$ bond currents, which connect N atoms in bulk InP. Left: bond currents relevant for $I_x^x$. Right: bond currents relevant for $I_y^x$. Bond currents represented in different colors have different magnitudes. }
\label{Fig_InN_bond_currents}
\end{figure}

\section{Bond currents in bulk InN}\label{App.Bond_InN}
InN has a wurtzite crystal structure. Some components of the ESC pseudotensor per unit cell do not vanish, as discussed at the end of Sec. \ref{sec.InP}. 
Here we show that these results can be understood by analysing the bond spin currents.\\
The largest bond currents in InN are between N atoms, and not between In atoms. 
This is similar to what found for InP, where the largest bond currents were those connecting the P atoms (see Sec. \ref{App.Bond_InP}). The $x$-spin bond currents in the rectangular unit cell of InP are presented in Fig. \ref{Fig_InN_bond_currents}. 
There are four N atoms in the cell. N$_1$ and N$_2$  are on the same $zy$ plane, while N$_3$ and N$_4$ are shifted along $x$ by half the lattice constant.
N$_1$ and N$_3$ are related to N$_2$ and N$_4$ via a trans-reflection, whose glide plane is parallel to $xz$. 
This symmetry implies that $\mathcal{I}^x_{13}=-\mathcal{I}^x_{24}$ and these two bond currents cancel each other along $x$. 
Similarly one can see that $\mathcal{I}^x_{14}=-\mathcal{I}^x_{23}$, so that they also cancel each other out along $x$. 
The other smaller bond currents follow the same symmetries and undergo identical cancellations. 
As a result, the global ESC component $I_x^x$ vanishes.\\
In contrast, $\mathcal{I}^x_{13}$ and $\mathcal{I}^x_{42}$ do not compensate each other along $y$, but they effectively add up. 
There is another large bond current in the supercell, namely $\mathcal{I}^x_{12}$. 
This has opposite sign with respect to $\mathcal{I}^x_{13}$ and $\mathcal{I}^x_{42}$ along $y$. It is however not connected by symmetry to them. 
Hence, there is no exact cancellation and ultimately the ESC component $I_y^x$ is finite.\\

\begin{figure*}[ht]
\centering\includegraphics[width=1.0\textwidth,clip=true]{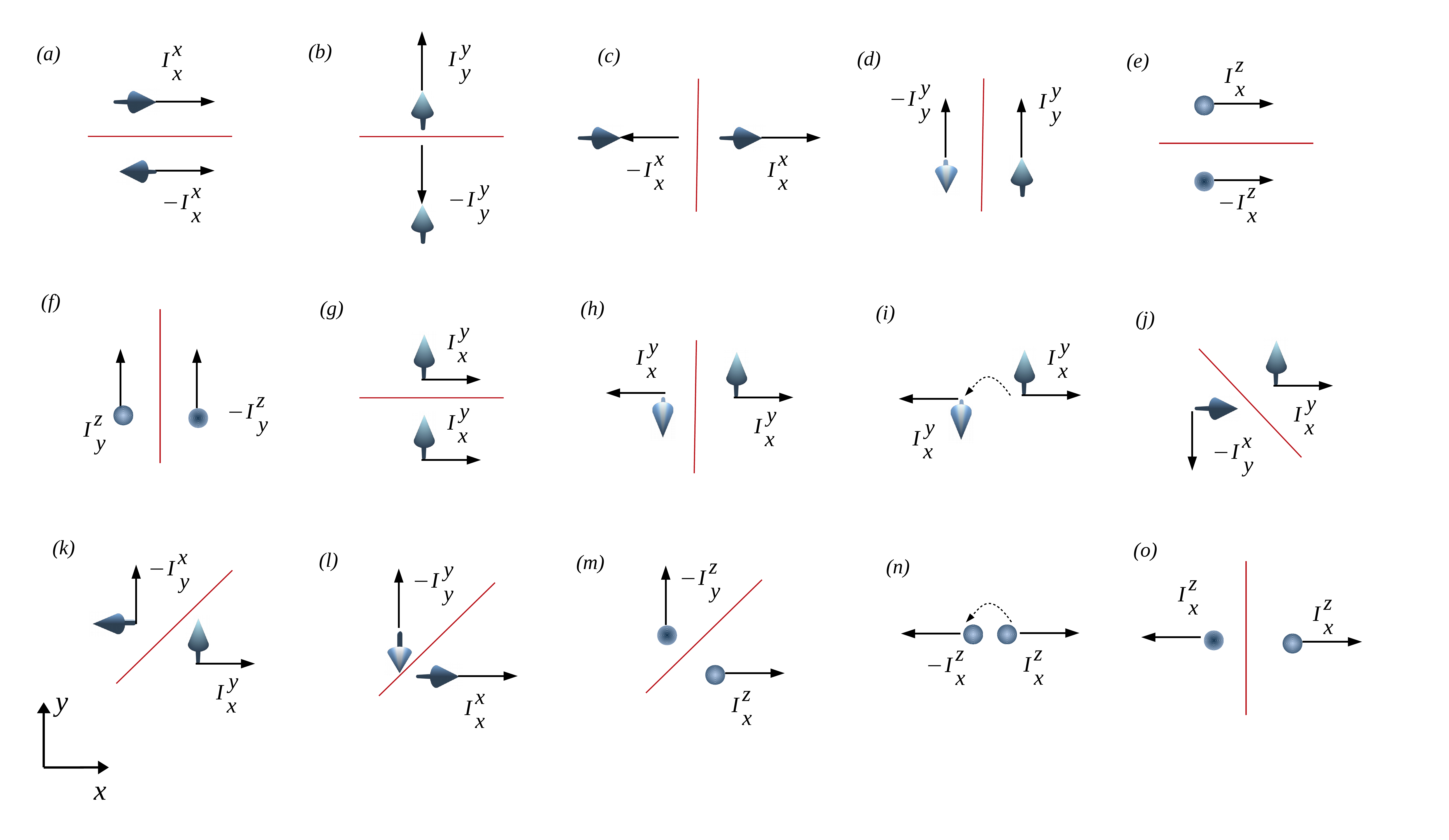}
\caption{Transformations of the momentum vector, of the spin pseudovector and therefore of the ESC components under various symmetry operations. 
The momentum and the spin are respectively represented as a black thin arrow and a rounded tridimensional arrow. Mirror reflection lines are painted in red. 180-degree rotations in (i) and (n) are represented as thin dashed lines terminating with an arrow.}
\label{fig.reflections}
\end{figure*}

\section{Symmetry analysis of the ESC pseudotensor}\label{App.Symmetry}
The components of the ESC pseudotensor in Eq. (\ref{ESCpseudo}) transform under symmetry operations as the direct product of the momentum vector and of the spin pseudovector. 
This is shown in Fig. \ref{fig.reflections}. Following Ref. \cite{Ganichev4}, we then use a simple general reasoning to derive the structure of the ESC pseudotensor for Au(001), Au(011), InP(001)- 
and InP(110)-oriented slabs.
\paragraph*{Au(011) surface.} The point symmetry group is $\mathbb{C}_{2v}$. It contains two mirror reflections through the lines $m_1$ and $m_2$, which are parallel to the $y$ and the $x$-axis, respectively [see Fig. \ref{fig.Au} (central panel)]. 
Additionally there is a 180-degree rotation around the normal axis. To obtain the structure of the ESC pseudotensor we need to determine which ones of its components remain invariant under these operations.\\
%We start by considering $I^x_{x}$ and $I^y_{y}$.
$I^x_{x}$ ($I^y_{y}$) transforms as the direct product
of the momentum and of the spin both parallel to the $x$- ($y$-)axis. The reflection through $m_2$ ($m_1$) leaves invariant the momentum vector, but it changes the direction of the spin pseudovector as shown in Fig. \ref{fig.reflections}-a(-d). Hence, $I^x_{x}$ ($I^y_{y}$) is reflected into $-I^x_{x}$ ($-I^y_{y}$). Both $I^x_{x}$ and $I^y_{y}$ vanish [the same conclusion can be reached by looking at  Fig. \ref{fig.reflections}-b(-c) instead of Fig. \ref{fig.reflections}-a(-d)].\\
%Next, we consider $-I^z_{x}$ and $I^z_{y}$. These components 
$I^z_{x}$ and $I^z_{y}$ correspond to the spin along the $z$-axis and the momentum along $x$ and $y$, respectively (Fig. \ref{fig.reflections}-e and -f). Using arguments similar to those above, we find that  $I^z_{x}$ ($I^z_{y}$) is reflected through $m_2$ ($m_1$) into $-I^z_{x}$ ($-I^z_{y}$). Thus, $I^z_{x}$ and $I^z_{y}$ are equal to zero.\\ 
$I^y_{x}$ transforms as the direct product of the
the momentum along the $x$-axis and the spin parallel to the $y$-axis. Neither the momentum nor the spin are affected by the reflection through $m_2$ (Fig. \ref{fig.reflections}-g), whereas both of them change sign after the reflection through $m_1$ (\ref{fig.reflections}-h). The net effect of this simultaneous sign change is nonetheless that $I^y_{x}$ remains invariant. $I^y_{x}$ is also left unchanged after the 180-degree rotation through the normal axis as such rotation flips both the momentum and the spin (Fig. \ref{fig.reflections}-i). The ESC component $I^y_{x}$ is therefore allowed by the $\mathbb{C}_{2v}$ symmetry. Similar arguments apply also for $I^x_{y}$, which is therefore allowed as well.\\
In conclusion $I^y_{x}$ and $I^x_{y}$ are the only non-zero components of the ESC pseudotensor, which has the same structure as in anisotropic Rashba systems \cite{Vajna} confirming the DFT results in the central panel of Fig. \ref{fig.Au}. \\
\paragraph*{Au(001) surface.} The point group is $\mathbb{C}_{4v}$. In addition to the mirror reflection lines $m_1$ and $m_2$, which are respectively parallel to the $y$- and $x$-axis, there are two other reflection lines, $m_3$ and $m_4$, along the diagonals (see the top panel of Fig. \ref{fig.Au}). Similarly to the Au(011) case, $I^x_{y}$ and $I^y_{x}$ are the only non-zero components of the ESC pseudotensor allowed by symmetry. 
However, we now have the additional relation $I^x_{y} = -I_{x}^y$ imposed by 
the reflection through $m_3$ and $m_4$. In fact, as shown in Fig. \ref{fig.reflections}-j, the momentum parallel to the $x$ ($y$) axis
is transformed into a negative momentum along $y$ ($x$) after the reflection through $m_4$. At the same time, the $x$- ($y$-)spin component is swapped with the $y$- ($x$-)spin component. 
Alternatively, one can reach the same results by analysing the reflection through $m_3$ (Fig. \ref{fig.reflections}-k). 
In conclusion, the ESC pseudotensor in Au(001) has the same structure as in isotropic Rashba systems \cite{Rashba}. 
This supports the DFT results in the top panel of Fig. \ref{fig.Au}.   \\

\begin{figure}[h]
\centering\includegraphics[width=0.4\textwidth,clip=true]{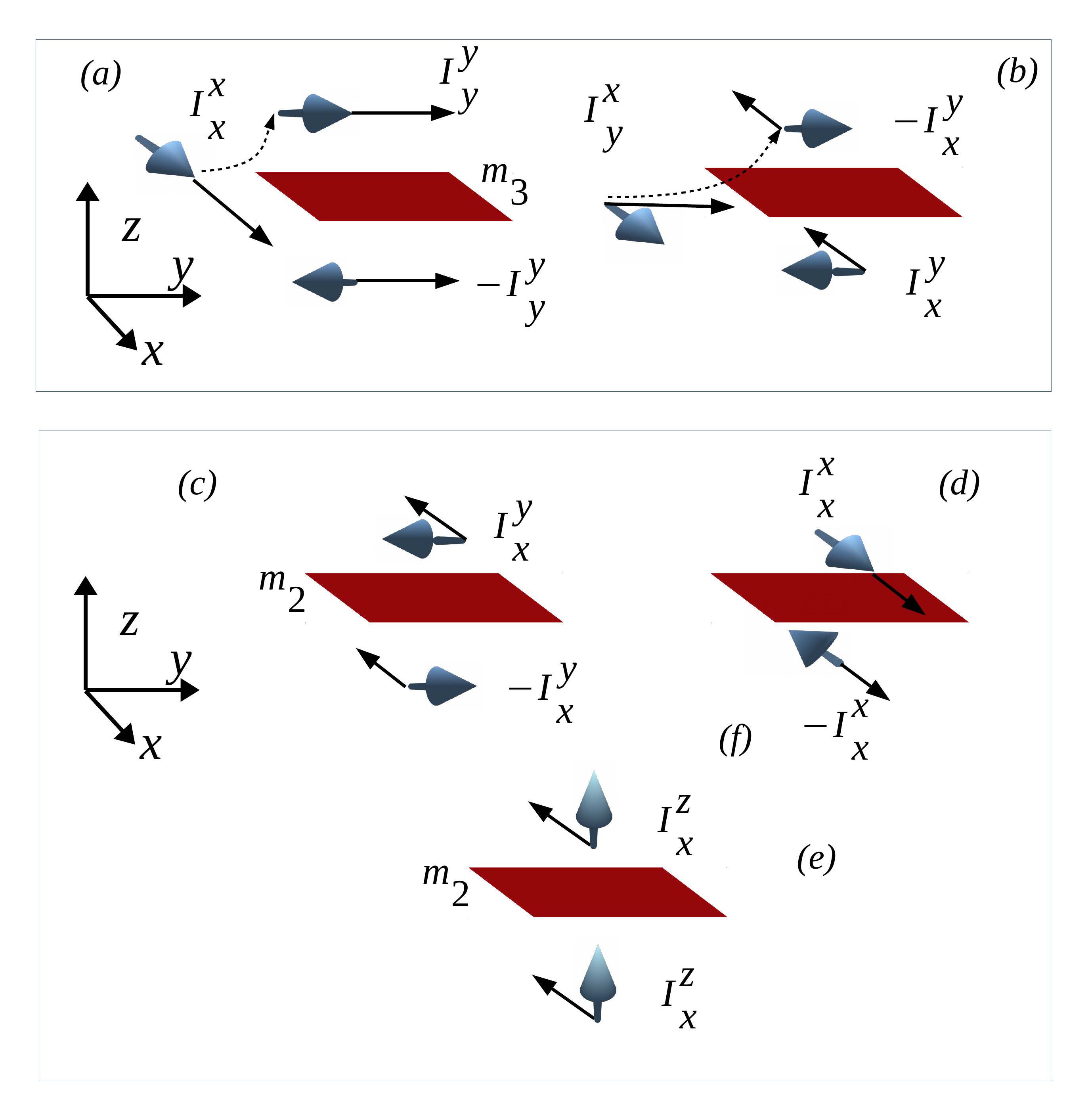}
\caption{Top-down symmetry transformations for the InP(001)-oriented slab (a and b) and the InP(011)-oriented slab (c, d and e).}
\label{fig.InP_100_sym}
\end{figure}

\paragraph*{InP$(001)$-oriented slab.} The studied slabs are presented in Fig. \ref{fig.InP_100}. The Cartesian $x$- and $y$-axis are parallel to the $(100)$ and $(010)$ directions, while the $z$-axis lays along the $(001)$ direction. 
The 21-layer and the 20-layer slabs have $\mathbb{D}_{2d}$ and $\mathbb{C}_{2v}$ point groups, which comprise a 180-degree rotation around a $C$ axis parallel to $z$ and two reflection planes, $m_1$ and $m_2$, which are respectively along the $(110)$ and the $(1\bar{1}0)$ directions. 
Additionally, $\mathbb{D}_{2d}$ has a top-down symmetry operation $S_4$, which is a $90$-degree rotation around the $C$ axis, followed by a reflection through the $m_3$ plane cutting the slab in two halves. \\
At each surface, the effect of the reflections through $m_1$ and $m_2$ are analysed using similar arguments as in the case of Au. In particular, we see that $I^x_{x,\mathrm{TS/BS}}$ ($I^y_{y,\mathrm{TS/BS}}$) transforms into $-I^y_{y,\mathrm{TS/BS}}$ ($-I^x_{x,\mathrm{TS/BS}}$) upon reflection through $m_1$ or $m_2$ (Fig. \ref{fig.reflections}-l). This operation also imposes that $I^x_{y,\mathrm{TS/BS}}$ and $I^y_{x,\mathrm{TS/BS}}$ are respectively equal to $-I^y_{x,\mathrm{TS/BS}}$ and $-I^x_{y,\mathrm{TS/BS}}$ (Fig. \ref{fig.reflections}-j and -k). These exact relations are fulfilled within numerical accuracy by our DFT results in Fig. \ref{fig.InP_100_layer}. \\
In the case of the 21-layer slab with $\mathbb{D}_{2d}$ point group, the transformation of the ESC components according to the roto-reflection $S_4$ is presented in Fig. \ref{fig.InP_100_sym}. 
We distinguish two cases.\\
\begin{itemize}
\item[1)] Both the momentum and the spin at the top surface point along the same direction, for instance $x$.
They are initially rotated to the $y$ direction, so that $I^x_{x,\mathrm{TS}}$ is changed into $I^y_{y,\mathrm{TS}}$ (\ref{fig.InP_100_sym}-a). Then, the top-down reflection converts the sign of the spin, but not that of the momentum, giving $I^x_{x,\mathrm{BS}}=-I^y_{y,\mathrm{TS}}$ (Fig. \ref{fig.InP_100_sym}-a). 
%The rotation transforms $I^x_{x,\mathrm{TS}}$ into $I^y_{y,\mathrm{TS}}$ and the top-down reflection finally gives $I^x_{x,\mathrm{BS}}=-I^y_{y,\mathrm{TS}}$ (Fig. \ref{fig.InP_100_sym}-e). 
Since $I^y_{y,\mathrm{TS}}=-I^x_{x,\mathrm{TS}}$ (Fig. \ref{fig.reflections}-l), we finally find $I^x_{x,\mathrm{BS}}= I^x_{x,\mathrm{TS}}$ and therefore $I^x_{x,\mathrm{slab}}=2I^x_{x,\mathrm{TS}}$. 
This is the result described in the main text.  \\
\item[2)] The momentum and the spin at the top surface are perpendicular. For example, we assume the momentum to lay along the $y$-axis and the spin along $x$-axis (Fig. \ref{fig.InP_100_sym}-b). The rotation followed by the top-down reflection gives $I^x_{y,\mathrm{TS}}=I^y_{x,\mathrm{BS}}$. Since $I^y_{x,\mathrm{TS}}=-I^x_{y,\mathrm{TS}}$ (Fig. \ref{fig.reflections}-j and -k), we finally have that
\begin{equation}
I^y_{x,\mathrm{slab}}=I^y_{x,\mathrm{TS}}+I^y_{x,\mathrm{BS}}=I^y_{x,\mathrm{TS}}-I^y_{x,\mathrm{TS}}=0,
\end{equation}
in agreement with the DFT results in the top panel of Fig. \ref{fig.InP_100}.\\
\end{itemize}
The $S_4$ roto-reflection is absent in the $\mathbb{C}_{2v}$ point group of the 20-layer slab.
The components of the ESCs at the two surfaces are therefore not related by symmetry.  $I^x_{y,\mathrm{slab}}$ and $I^y_{x,\mathrm{slab}}$ assume non-zero values because there is no cancellation between the ESCs at top and bottom surfaces. This confirms the structure of the ESC pseudotensor calculated by DFT and presented in the bottom panel of Fig. \ref{fig.InP_100}.\\
Finally, we analyse the current for the $z$ spin component. The reflection through $m_1$ and $m_2$ implies that $I^z_{x,\mathrm{slab}}=-I^z_{y,\mathrm{slab}}$ (Fig. \ref{fig.reflections}-m). However, both $\mathbb{D}_{2d}$ and $\mathbb{C}_{2v}$ contain a 180-degree rotation around the $C$ axis normal to the slab surfaces. As seen in Fig. \ref{fig.reflections}-n this transforms $I^z_{x,\mathrm{slab}}$ ($I^z_{y,\mathrm{slab}}$) into $-I^z_{x,\mathrm{slab}}$ ($-I^z_{y,\mathrm{slab}}$) and, as a result, $I^z_{x,\mathrm{slab}}=0$ ($I^z_{y,\mathrm{slab}}=0$). Hence, all elements in the third raw of the ESC pseudotensor are zero as seen in Fig. \ref{fig.InP_100}.     \\ 
 
\paragraph*{InP$(011)$-oriented slab.}The studied slab is presented in Fig. \ref{fig.InP_110}.
The Cartesian $x$- and $y$-axis lay parallel to the $(101)$ and $(\bar 110)$ directions, respectively. The slab has point group $\mathbb{C}_{2v}$. There are therefore two mirror reflection planes $m_1$ and $m_2$. $m_1$ is parallel to the $yz$ plane, whereas $m_2$ is associated to the top-down symmetry of the slab and is parallel to the $xy$ plane.\\
The ESC components $I^x_{x,\mathrm{slab}}$ and $I^y_{y,\mathrm{slab}}$ are transformed into $-I^x_{x,\mathrm{slab}}$ and $-I^y_{y,\mathrm{slab}}$ after reflection through $m_1$ (Fig. \ref{fig.reflections}-c and -d). 
The component $I^z_{y,\mathrm{slab}}$ changes sign in a similar fashion (Fig. \ref{fig.reflections}-f). Only $I^x_{y,\mathrm{slab}}$, $I^y_{x,\mathrm{slab}}$ and $I^z_{x,\mathrm{slab}}$ are invariant at the reflection through $m_1$ (Fig. \ref{fig.reflections}-h and -o). 
However, the reflection through the second mirror plane $m_2$ transforms $I^x_{y,\mathrm{slab}}$ and $I^y_{x,\mathrm{slab}}$ into $-I^x_{y,\mathrm{slab}}$ and $-I^y_{x,\mathrm{slab}}$ (Fig. \ref{fig.InP_100_sym}-c). 
Thus the only component, which is left invariant, is $I^z_{x,\mathrm{slab}}$ (Fig. \ref{fig.InP_100_sym}-e). This result explains the structure of the ESC pseudotensor in Fig. \ref{fig.InP_110}.\\

\end{document}